\documentclass[journal,twoside]{IEEEtran}
\usepackage{amsmath,amsfonts}
\usepackage{amssymb}
\usepackage{fancyvrb}
\usepackage{algorithmic}
\usepackage[linesnumbered,lined,ruled,vlined]{algorithm2e}
\usepackage{array}
\usepackage{textcomp}
\usepackage{stfloats}
\usepackage{url}
\usepackage{bm}
\usepackage{verbatim}
\usepackage{graphicx}
\usepackage[usenames,dvipsnames]{xcolor}
\usepackage{tikz}
\usetikzlibrary{spy}
\usepackage{pgfplots}
\usepackage{cite}
\usepackage{booktabs}
\usepackage{threeparttable}
\usepackage{makecell}
\usepackage{multirow}
\usepackage{circledsteps}
\usepackage{enumitem}
\usepackage{pifont}
\definecolor{bostonunired}{HTML}{CC0000}    % https://www.color-name.com/hex/CC0000
\definecolor{fireenginered}{HTML}{CF202A}   % https://www.color-name.com/hex/CF202A
\definecolor{vermillionred}{HTML}{E34234}   % https://www.color-name.com/vermillion.color

\definecolor{titlered}{RGB}{212,0,0}

\definecolor{reddeep}{RGB}{178,24,43}
\definecolor{redlight}{RGB}{252,78,42}
\definecolor{redlight2}{RGB}{255,114,111}

\definecolor{redbright}{RGB}{255,24,43}

\definecolor{perfectgreen}{HTML}{4FBF26}    % https://www.color-name.com/perfect-green.color
\definecolor{maygreen}{HTML}{4E9B47}        % https://www.color-name.com/hex/4E9B47
\definecolor{mattelime}{HTML}{75AD50}       % https://www.color-name.com/matte-lime.color

\definecolor{indigorainbow}{HTML}{1D3F6E}   % https://www.color-name.com/hex/1D3F6E
\definecolor{royalazure}{HTML}{1866E1}      % https://www.color-name.com/royal-azure.color
\definecolor{azure}{HTML}{008AFF}           % https://www.color-name.com/azure-traditional.color
\definecolor{blueberry}{HTML}{4F86F7}       % https://www.color-name.com/blueberry.color
\definecolor{fadednavy}{HTML}{242F78}       % https://www.color-name.com/faded-navy.color
\definecolor{navy}{HTML}{000080}            % https://www.color-name.com/navy.color

\definecolor{goldwebgolden}{HTML}{FFD700}   % https://www.color-name.com/hex/FFD700
\definecolor{radioactive}{HTML}{FAE500}     % https://www.color-name.com/radioactive.color
\definecolor{gold}{HTML}{FFD700}            % https://www.color-name.com/gold.color

\definecolor{vividgamboge}{HTML}{FF9900} % https://www.color-name.com/hex/FF9900

\definecolor{shinygray}{HTML}{C7C6C6}

\definecolor{slategrey}{HTML}{708090}   % https://www.color-name.com/slate-gray.color
\definecolor{neutralgray}{HTML}{828382} % https://www.color-name.com/neutral-grey.color

\definecolor{mattecharcoal}{HTML}{3B4248} % https://www.color-name.com/matte-charcoal.color

\definecolor{graydark}{HTML}{6B6E70} % https://www.color-name.com/antique-steel.color

\definecolor{charcoal}{HTML}{36454F} % https://www.color-name.com/charcoal.color
\definecolor{charcoalShade1}{HTML}{2B373F}
\definecolor{charcoalShade2}{HTML}{263037}
\definecolor{charcoalShade3}{HTML}{20292F}

\definecolor{sunset1}{HTML}{F3E79B}
\definecolor{sunset2}{HTML}{FAC484}
\definecolor{sunset3}{HTML}{F8A07E}
\definecolor{sunset4}{HTML}{EB7F86}
\definecolor{sunset5}{HTML}{CE6693}
\definecolor{sunset6}{HTML}{A059A0}
\definecolor{sunset7}{HTML}{5C53A5}

\definecolor{bluyi1}{HTML}{F7FEAE}
\definecolor{bluyi2}{HTML}{B7E6A5}
\definecolor{bluyi3}{HTML}{7CCBA2}
\definecolor{bluyi4}{HTML}{46AEA0}
\definecolor{bluyi5}{HTML}{089099}
\definecolor{bluyi6}{HTML}{00718B}
\definecolor{bluyi7}{HTML}{045275}

\definecolor{geyser1}{HTML}{008080}
\definecolor{geyser2}{HTML}{70A494}
\definecolor{geyser3}{HTML}{B4C8A8}
\definecolor{geyser4}{HTML}{F6EDBD}
\definecolor{geyser5}{HTML}{EDBB8A}
\definecolor{geyser6}{HTML}{DE8A5A}
\definecolor{geyser7}{HTML}{CA562C}

\definecolor{temps1}{HTML}{009392}
\definecolor{temps2}{HTML}{39B185}
\definecolor{temps3}{HTML}{9CCb86}
\definecolor{temps4}{HTML}{E9E29C}
\definecolor{temps5}{HTML}{EEB479}
\definecolor{temps6}{HTML}{E88471}
\definecolor{temps7}{HTML}{CF597E}

\definecolor{earth1}{HTML}{A16928}
\definecolor{earth2}{HTML}{BD925A}
\definecolor{earth3}{HTML}{D6BD8D}
\definecolor{earth4}{HTML}{EDEAC2}
\definecolor{earth5}{HTML}{B5C8B8}
\definecolor{earth6}{HTML}{79A7AC}
\definecolor{earth7}{HTML}{2887A1}

\definecolor{bold1}{HTML}{7F3C8D}
\definecolor{bold2}{HTML}{11A579}
\definecolor{bold3}{HTML}{3969AC}
\definecolor{bold4}{HTML}{F2B701}
\definecolor{bold5}{HTML}{E73F74}
\definecolor{bold6}{HTML}{80BA5A}
\definecolor{bold7}{HTML}{E68310}
\definecolor{bold8}{HTML}{008695}
\definecolor{bold9}{HTML}{CF1C90}
\definecolor{bold10}{HTML}{F97B72}
\definecolor{bold11}{HTML}{4B4B8F}
\definecolor{bold12}{HTML}{A5AA99}

\definecolor{pastel1}{HTML}{66C5CC}
\definecolor{pastel2}{HTML}{F6CF71}
\definecolor{pastel3}{HTML}{F89C74}
\definecolor{pastel4}{HTML}{DCB0F2}
\definecolor{pastel5}{HTML}{87C55F}
\definecolor{pastel6}{HTML}{9EB9F3}
\definecolor{pastel7}{HTML}{FE88B1}
\definecolor{pastel8}{HTML}{C9DB74}
\definecolor{pastel9}{HTML}{8BE0A4}
\definecolor{pastel10}{HTML}{B497E7}
\definecolor{pastel11}{HTML}{D3B484}
\definecolor{pastel12}{HTML}{B3B3B3}

\definecolor{prism1}{HTML}{5F4690}
\definecolor{prism2}{HTML}{1D6996}
\definecolor{prism3}{HTML}{38A6A5}
\definecolor{prism4}{HTML}{0F8554}
\definecolor{prism5}{HTML}{73AF48}
\definecolor{prism6}{HTML}{EDAD08}
\definecolor{prism7}{HTML}{E17C05}
\definecolor{prism8}{HTML}{CC503E}
\definecolor{prism9}{HTML}{94346E}
\definecolor{prism10}{HTML}{6F4070}
\definecolor{prism11}{HTML}{994E95}
\definecolor{prism12}{HTML}{666666}

\definecolor{vivid1}{HTML}{E58606}
\definecolor{vivid2}{HTML}{5D69B1}
\definecolor{vivid3}{HTML}{52BCA3}
\definecolor{vivid4}{HTML}{99C945}
\definecolor{vivid5}{HTML}{CC61B0}
\definecolor{vivid6}{HTML}{24796C}
\definecolor{vivid7}{HTML}{DAA51B}
\definecolor{vivid8}{HTML}{2F8AC4}
\definecolor{vivid9}{HTML}{764E9F}
\definecolor{vivid10}{HTML}{ED645A}
\definecolor{vivid11}{HTML}{CC3A8E}
\definecolor{vivid12}{HTML}{A5AA99}

\definecolor{cBrewerPaired1}{HTML}{A6CEE3}
\definecolor{cBrewerPaired2}{HTML}{1F78B4}
\definecolor{cBrewerPaired3}{HTML}{B2DF8A}
\definecolor{cBrewerPaired4}{HTML}{33A02C}
\definecolor{cBrewerPaired5}{HTML}{FB9A99}
\definecolor{cBrewerPaired6}{HTML}{E31A1C}
\definecolor{cBrewerPaired7}{HTML}{FDBF6F}
\definecolor{cBrewerPaired8}{HTML}{FF7F00}
\definecolor{cBrewerPaired9}{HTML}{CAB2D6}
\definecolor{cBrewerPaired10}{HTML}{6A3D9A}
\definecolor{cBrewerPaired11}{HTML}{FFFF99}
\definecolor{cBrewerPaired12}{HTML}{B15928}

\definecolor{cBrewerQualPrint1}{HTML}{E41A1C}
\definecolor{cBrewerQualPrint2}{HTML}{377EB8}
\definecolor{cBrewerQualPrint3}{HTML}{4DAF4A}
\definecolor{cBrewerQualPrint4}{HTML}{984EA3}
\definecolor{cBrewerQualPrint5}{HTML}{FF7F00}
\definecolor{cBrewerQualPrint6}{HTML}{FFFF33}
\definecolor{cBrewerQualPrint7}{HTML}{A65628}
\definecolor{cBrewerQualPrint8}{HTML}{F781BF}
\definecolor{cBrewerQualPrint9}{HTML}{999999}

\definecolor{thsViolet1}{HTML}{E4C7F1} % Purp 2
\definecolor{thsViolet2}{HTML}{9F82CE} % Purp 5
\definecolor{thsViolet3}{HTML}{4B4B8F} % Bold 11

\definecolor{thsPink1}{HTML}{DCB0F2} % Pstel 4
\definecolor{thsPink2}{HTML}{FE88B1} % Pastel 7
\definecolor{thsPink3}{HTML}{CC3A8E} % Vivid 11

\definecolor{thsBlue1}{HTML}{4F86F7} % https://www.color-name.com/blueberry.color
\definecolor{thsBlue2}{HTML}{0F52BA} % https://www.color-name.com/sapphire.color
\definecolor{thsBlue3}{HTML}{332288} % Safe 5

\definecolor{thsCyan1}{HTML}{88CCEE} % Safe 1
\definecolor{thsCyan2}{HTML}{66C5CC} % Pastel 1
\definecolor{thsCyan3}{HTML}{367588} % https://www.color-name.com/teal-blue.color

\definecolor{thsGreen1}{HTML}{4FBF26} % https://www.color-name.com/perfect-green.color
\definecolor{thsGreen2}{HTML}{11A579} % Bold 2
\definecolor{thsGreen3}{HTML}{0F8554} % Prism 4

\definecolor{thsYellow1}{HTML}{FFFF66} % https://www.color-name.com/laser-lemon.color
\definecolor{thsYellow2}{HTML}{FAE500} % https://www.color-name.com/radioactive.color
\definecolor{thsYellow2}{HTML}{FFFF00}

\definecolor{thsOrange1}{HTML}{ECDA9A} % OrYel 1
\definecolor{thsOrange2}{HTML}{F7945D} % OrYel 3
\definecolor{thsOrange3}{HTML}{E58606} % Vivid 1

\definecolor{thsRed1}{HTML}{CC503E} % Prism 8
\definecolor{thsRed2}{HTML}{CC0000} % https://www.color-name.com/hex/CC0000
\definecolor{thsRed3}{HTML}{670E10} % https://www.color-name.com/royal-maroon.color

\definecolor{thsGray1}{HTML}{B1B6B7} % https://www.color-name.com/antique-steel.color
\definecolor{thsGray2}{HTML}{828382} % https://www.color-name.com/neutral-grey.color
\definecolor{thsGray3}{HTML}{36454F} % https://www.color-name.com/charcoal.color

\def\RZeroColor{white}
\def\ROneColor{black}

\definecolor{myblue}{HTML}{BDD7EE}
\definecolor{mygreen}{HTML}{C5E0B4}
\definecolor{myyellow}{HTML}{FFF2CC}
\def\mySpcColor{myyellow}
\def\myRepColor{mygreen}
\def\mySRColor{myblue}

\definecolor{barGREEN}{HTML}{A9D18E}
\definecolor{barYELLOW}{HTML}{FFD966}
\definecolor{barBLUE}{HTML}{2E75B6}
\definecolor{barGRAY}{HTML}{BFBFBF}
\definecolor{barRED}{HTML}{FF0000}

\definecolor{barGREEN2}{HTML}{13B062}%{129E63}

\definecolor{TypeIIIred}{HTML}{F4B183}
\def\TYPEIIIColor{TypeIIIred}

\newcommand{\tikzcircle}[2][black,fill=black]{\protect\tikz[baseline=-0.8ex]\protect\draw[#1,radius=#2] (0,0) circle ;}%
\newcommand{\tikzcirclewidth}{0.15}
\newcommand*\myscale{1}
\newcommand*\circled[2][1]{\renewcommand*\myscale{#1}\tikz[baseline=(char.base)]{
            \node[shape=circle,draw,inner sep=1.5pt, scale=\myscale] (char) {#2};}}

\newcommand{\eqrangelabel}[2]{(\ref{#1}--\ref{#2})}

\newcommand{\mytextsf}[1]{{\small\textsf{#1}}}
\newcommand{\mc}[3]{\multicolumn{#1}{#2}{#3}}

\DeclareMathOperator*{\OFG}{\pi_{\ast}}
\DeclareMathOperator*{\sgn}{sgn}
\DeclareMathOperator*{\FrameX}{\mytextsf{Frame~0}}
\DeclareMathOperator*{\FrameY}{\mytextsf{Frame~1}}
\DeclareMathOperator*{\MemX}{\mytextsf{Mem~0}}
\DeclareMathOperator*{\MemY}{\mytextsf{Mem~1}}
\DeclareMathOperator*{\SX}{\mathcal{S}_1}
\DeclareMathOperator*{\SY}{\mathcal{S}_2}

\begin{document}

\title{A Node-Based Polar List Decoder with Frame Interleaving and Ensemble Decoding Support}

\author{Yuqing~Ren,
Leyu~Zhang,
Ludovic~Damien~Blanc,
Yifei~Shen,
Xinwei~Li,\\
Alexios~Balatsoukas-Stimming,
Chuan~Zhang,
Andreas~Burg
\vspace{-0.2cm}
\thanks{Y. Ren, L. D. Blanc, Y. Shen, X. Li and A. Burg are with the Telecommunications Circuits Laboratory, \'{E}cole Polytechnique F\'{e}d\'{e}rale de Lausanne (EPFL), Laussane 1015, Switzerland.~\emph{(Yuqing Ren and Leyu Zhang equally contributed to this work. Corresponding author: Andreas Burg)}}
\thanks{L. Zhang is with the Telecommunications Circuits Laboratory, \'{E}cole Polytechnique F\'{e}d\'{e}rale de Lausanne (EPFL), Laussane 1015, Switzerland and also with the CAS Key Laboratory of Wireless-Optical Communications, University of Science and Technology of China (USTC), China.}
\thanks{A. Balatsoukas-Stimming is with the Department of Electrical Engineering,
	Eindhoven University of Technology, 5600 MB Eindhoven, The Netherlands.}
\thanks{C. Zhang is with the Lab of Efficient Architectures for Digital-communication and Signal-processing (LEADS), Southeast University, China.}
}
\maketitle

\begin{abstract}
Node-based successive cancellation list (SCL) decoding has received considerable attention in wireless communications for its significant reduction in decoding latency, particularly with 5G New~Radio~(NR) polar codes.
However, the existing node-based SCL decoders are constrained by sequential processing, leading to complicated and data-dependent computational units that introduce unavoidable stalls, reducing hardware efficiency.
In this paper, we present a frame-interleaving hardware architecture for a generalized node-based SCL decoder.
By efficiently reusing otherwise idle computational units, two independent frames can be decoded simultaneously, resulting in a significant throughput gain.
Based on this new architecture, we further exploit graph ensembles to diversify the decoding space, thus enhancing the error-correcting performance with a limited list size.
Two dynamic strategies are proposed to eliminate the residual stalls in the decoding schedule, which eventually results in nearly $2\times$ throughput compared to the state-of-the-art baseline node-based SCL decoder.
To impart the decoder rate flexibility, we develop a novel online instruction generator to identify the generalized nodes and produce instructions on-the-fly.
The corresponding $28$nm FD-SOI ASIC SCL decoder with a list size of $8$ has a core area of $1.28$~mm$^2$ and operates at $692$~MHz.
It is compatible with all 5G~NR polar codes and achieves a throughput of $3.34$~Gbps and an area efficiency of $2.62$~Gbps/mm$^2$ for uplink $(1024,512)$ codes, which is $1.41\times$ and $1.69\times$ better than the state-of-the-art node-based SCL decoders.
\end{abstract}


\section{Introduction}
\IEEEPARstart{P}{olar} codes~\cite{arik:09} have been ratified as the standard codes to protect control channels in 5G enhanced mobile broadband (eMBB)~\cite{5G}, thus raising significant attention in both academia and industry.
Successive cancellation (SC)~\cite{arik:09} and SC list (SCL) decoding~\cite{Tal15List} led the mainstream development of polar decoding algorithms and implementations. 
In particular, SCL decoding can evaluate two hypotheses for each information bit, by estimating each information bit as either $0$ or $1$ and maintaining a list of up to $L$ candidate codewords in parallel during the decoding. 
This yields superior error-correcting performance compared to SC decoding.
Using hardware-friendly log-likelihood ratio (LLR)-based SCL decoding~\cite{Bala15LLR}, an SCL decoder with $L=8$ provides a good trade-off between error-correcting performance and hardware complexity and has been chosen as the algorithm baseline during the 5G standardization process~\cite{5GSCL}.

In terms of hardware implementation, to reduce long decoding latency in conventional SC-based decoders~\cite{Bala15LLR,giard2017polarbear}, node-based techniques have been introduced to SC decoders~\cite{Sarkis14Fast,Hanif17Fast,Condo18Generalized}, in which special bit patterns of constituent codes can be identified in the decoding tree to allow direct decoding of multiple bits in parallel instead of repeatedly traversing the corresponding parts of the tree.
Then, such techniques were also introduced to SCL decoders, using list extension strategies for special nodes~\cite{sarkis2016fast,Hashemi17Fast,Hanif18Fast}, thereby significantly decreasing the decoding latency.
Notably, the recently found generalized sequence repetition node (SR)~\cite{Zheng21Threshold} can support most existing node patterns in SC decoding.
The work in~\cite{Ren22Sequence} first extends the SR node to SCL decoding and implements a corresponding SR-List decoder to meet the stringent low-latency and low-complexity requirements of 5G.
However, the existing SCL decoders (including~\cite{Ren22Sequence}) still suffer from sequential processing in the special nodes, which leads to unavoidable stalls.
To mitigate the impact of these stalls, based on the architecture in~\cite{Hashemi17Fast}, several latency-optimization techniques have been proposed in~\cite{Kam23low} to reduce idle cycles by modifying the architectures of computational units to allow for~overlapped~processing.

In addition to low latency, high reliability is also essential, but the error-correcting capability of SCL decoding, with a limited list size, falls short in some resource-constrained scenarios and demands further enhancement.
Recently, automorphism ensemble decoding (AED) gained a lot of attention due to its superior performance and universality~\cite{Geiselhart19ISIT,geiselhart2021automorphism,Kestel2023SCC,johannsen2023successive}.
AED employs code automorphisms, represented as a block lower triangular affine (BLTA) group in polar codes~\cite{Geiselhart19ISIT}, to decode various codeword permutations in parallel, demonstrating particular effectiveness for short-length polar codes~\cite{Kestel2023SCC}.
However, its hard-wired routing of permutations lacks flexibility for diverse automorphisms.
It is noteworthy that as a subset of a general affine (GA) group, factor graph permutations~\cite{Elke18Belief,Ren2020TVT} (also known as permuted factor graphs, PFGs) efficiently widen the decoding space through graph ensemble decoding and can be implemented in hardware to generate these permuted graphs on-the-fly~\cite{Ren24BPL}.
However, this factor graph permutation alters the bit indices~\cite{Doan18Decoding} and thus greatly affects the node structure of the polar decoding tree.
This alteration complicates the direct application of node techniques to graph ensemble decoding.
Therefore, integrating graph ensembles with SCL decoding and node-based techniques for high performance and low latency remains challenging.

Moreover, these node-based polar decoders rely on the identification of the type and the length of special codes, which necessitates complex control logic~\cite{Sarkis14Fast}.
To bypass this issue, most decoders generate instructions offline and store the resulting instructions in dedicated memories.
The decoder then proceeds by reading these instructions to coordinate each component.
However, this instruction storage incurs significant area overhead~\cite{Sarkis14Fast,hashemi2016fast,Hashemi17Fast}.
Given the wide range of message lengths from $12$ to $1706$ in 5G~NR polar codes, it is impractical to store all possibilities and a rapid succession of code changes leaves no time to generate instructions.
To overcome this issue, an online generator is proposed in~\cite{hashemi2019rate} to identify the special patterns of basic nodes directly in hardware based on the channel relative reliability vector, thus supporting a rate-flexible decoder.
However, to the best knowledge of the authors, there is currently no online generator compatible with more generalized nodes, such as SR nodes, that achieves both strict low decoding latency and rate flexibility in generalized node-based SCL decoders.

\subsection*{Contributions:}
This work is an extension of our work in~\cite{Zhang2024ISCAS}, in which the node-based SCL decoder with frame-interleaving was initially proposed.
However, due to the mismatch in the processing cycles of computational units, some inevitable stalls still exist and they degrade throughput and area efficiency improvements~\cite{Zhang2024ISCAS}.
Moreover, the decoder from~\cite{Zhang2024ISCAS} relies on storing offline-generated instructions, which compromises its rate flexibility.
In this paper, we enhance the generalized node-based SCL decoder with frame-interleaving to address the above issues while remaining compatible with all 5G~NR polar codes.
Our contributions comprise the~following:
\begin{enumerate}
\setlength{\itemsep}{0pt}
\setlength{\parsep}{0pt}
\setlength{\parskip}{0pt}
\item
We propose a node-based SCL decoder with frame-interleaving, based on which, two independent frames can be decoded simultaneously by reusing the computational units in an alternating fashion.
This architecture significantly enhances the throughput at the cost of a small area increment.
Moreover, our decoder has three working modes:~\emph{Mode-I Frame-Interleaving} tailored to low latency,~\emph{Mode-II Graph-Interleaving} tailored to high performance, and~\emph{Mode-III Hybrid-Interleaving} tailored to high~efficiency.
\item
To effectively eliminate the residual stalls in the interleaving schedule, we provide two dynamic strategies: overlapped processing of SR nodes and dynamic path forking for high-rate nodes.
Based on these two strategies, our decoder can ultimately reduce $49\%$ of decoding cycles (the ideal is $50\%$) with negligible to no performance degradation.
\item
We design an online instruction generator that allows the identification of SR nodes from the binary sequence that specifies the information set and generates the list of decoder instructions on-the-fly.
Compared to a straightforward instruction memory, our online instruction generator yields a $97.5\%$ smaller area even for a single code and imparts the proposed SCL decoder~rate~flexibility.
\item
We implement the above frame-interleaving node-based SCL decoder, compatible with 5G~NR polar codes. 
The $28$nm FD-SOI post-layout implementation has a core area of $1.28$~mm$^{2}$, achieves a throughput of $3.34$~Gbps at $692$~MHz, and has an energy consumption of $60.32$~pJ/bit with a supply voltage of $1.0$~V.

\end{enumerate}

The remainder of this paper is organized as follows: 
Section~\ref{sec:pre} provides symbol definitions and background on polar codes and decoding. 
Section~\ref{sec:hardware} introduces the proposed architecture with frame-interleaving and three working modes. 
In Section~\ref{sec:strategy}, we present two optimization schemes to further reduce the latency. 
Section~\ref{sec:instruction} presents a novel online instruction generator which is compatible with basic nodes and SR nodes.
Section~\ref{sec:implem} discusses the implementation results and compares them with other works. 
Section~\ref{sec:conclusion} concludes~the~paper.

\section{Preliminaries}\label{sec:pre}
\emph{Notation:} Throughout this paper, we follow the definitions introduced below.
Boldface lowercase letters $\bm{u}$ denote vectors, where $\bm{u}[i]$ refers to the $i$-th element of $\bm{u}$ and $\bm{u}[i:j]$ is the sub-vector $(\bm{u}[i], \bm{u}[i+1],\dots, \bm{u}[j]),i\leq j$ and the null vector otherwise.
Boldface uppercase letters $\mathbf{B}$ represent matrices, where $\mathbf{B}[i][j]$ denotes the element at the $i$-th row and $j$-th column of $\mathbf{B}$.
Blackboard letters $\mathbb{S}$ mean sets with $|\mathbb{S}|$ being the cardinality.
We adopt the following parameters from the 5G~NR standard~\cite{5Gstandard}: $N=2^{n}$ is the length of the mother polar code, $K$ is the number of message bits, $K^\prime=K+P$ is the number of information bits with length-$P$ CRC bits, and $E$ the length of the codeword after rate-matching.
We denote a 5G~NR polar code as $(E,K)$, and the corresponding frozen and information sets are defined as~$\mathbb{F}$ and~$\mathbb{A}$.
For brevity, we use uplink~(UL) and downlink~(DL) polar codes from 5G~NR.
More detailed formulation of the information set and rate-matching principles for 5G~NR polar codes are described in~\cite{5G}.
We use one \emph{frame} to describe one codeword during the transmission.
Note that all indices regarding decoding~start~from~$0$.

\subsection{Construction and Encoding}
Given a length-$N$ input bit sequence $\bm{u}$, the encoded vector $\bm{x}$ is generated by $\bm{x}=\bm{u}\cdot\mathbf{G}$, where $\mathbf{G}=\mathbf{F}^{\otimes n}$ is the $n$-fold Kronecker product of $\mathbf{F}=\footnotesize{\begin{bmatrix} 1 & 0 \\ 1 & 1 \end{bmatrix}}$.
Based on the principle of channel polarization~\cite{arik:09}, the $K^\prime$ most reliable bit channels transmit information bits along with CRC bits, while the remaining $N-K^\prime$ bit channels transmit frozen bits, typically set to a value of $0$.
The $\mathbf{G}$-based factor graph of length-$N$ polar codes is called the original factor graph (OFG), defined as $\OFG$ including $n+1$ stages.
Given a graph permutation, we represent any PFG as $\pi$, and the design space is $\bm{\pi}$ which contains $|\bm{\pi}|=n!-1$ possible~PFGs.

\subsection{SC and SCL Decoding}
SC decoding can be represented as the traversal over a binary decoding tree with $n+1$ stages, where SC decoding successively updates the bit likelihoods.
Let $\mathcal{N}_{s,i}$ denote the $i$-th node (the node length is $2^{s}$) at stage $s$, where $0\leq i<2^{n-s}$ and $0\leq s\leq n$.
For $\mathcal{N}_{s,i}$, a length-$2^{s}$ LLR vector $\bm{\lambda}_{s,i}$ is received and after traversing all the child nodes, the node returns a length-$2^{s}$ partial sum (PSUM) vector $\bm{\beta}_{s,i}$.
The detailed update rules are described in~\eqref{eq:llr_fg_func}
\begin{equation}\label{eq:llr_fg_func}
\begin{aligned}
	\bm{\lambda}_{s,2i}[j]   &= f(\bm{\lambda}_{s+1,i}[j],\bm{\lambda}_{s+1,i}[j+2^s]) \,,\\
	\bm{\lambda}_{s,2i+1}[j] &= g(\bm{\lambda}_{s+1,i}[j],\bm{\lambda}_{s+1,i}[j+2^s],\bm{\beta}_{s,2i}[j]) \,,
\end{aligned}
\end{equation}
where $f$- and $g$-functions are defined in~\eqref{eq:fg_func_def}.
\begin{equation}\label{eq:fg_func_def}
\begin{aligned}
	f(x,y)  & \approx \sgn(x) \sgn(y) \min\{|x|,|y|\} \,, \\
	g(x,y,z) &= (1-2z) x + y  \,.
\end{aligned}
\end{equation}
The update rule for PSUM is shown in~\eqref{eq:psum_func}.
\begin{equation}\label{eq:psum_func}
\begin{aligned}
	\bm{\beta}_{s+1,i}[j]     &= \bm{\beta}_{s,2i}[j]\oplus\bm{\beta}_{s,2i+1}[j] \,,\\
	\bm{\beta}_{s+1,i}[j+2^s] &= \bm{\beta}_{s,2i+1}[j] \,.
\end{aligned}
\end{equation}
To mitigate the serial error propagation caused by bit errors in SC decoding, SCL decoding applies a breadth-first search to identify promising candidate codewords in the decoding tree.
At the cost of higher storage and computing complexity to enhance reliability, SCL decoding maintains a list of up to $L$ paths by examining both hypotheses for each information bit.
The reliabilities of the path candidates are measured by a path metric (PM)~\cite{Bala15LLR}, which is calculated~as
\begin{equation}\label{eq:PM}
\mathrm{PM}_i^{(l)}\approx\left\{
\begin{aligned}
	&\mathrm{PM}_{i-1}^{(l)}+|\lambda_{0,i}^{(l)}|, \qquad&&\text{if}\;\hat{\bm{u}}[i]\neq\mathrm{HD}(\lambda_{0,i}^{(l)}),\\
	&\mathrm{PM}_{i-1}^{(l)}, \qquad&&\text{otherwise}.
\end{aligned}
\right.
\end{equation}
Here $\mathrm{PM}_{i}^{(l)}$ denotes the PM of the $l$-th path after obtaining the estimated value $\hat{\bm{u}}[i]$ of the $i$-th bit, where $0\leq l<L$, $0\leq i<N$.
The hard decision function is defined as $\mathrm{HD}(x)=1$ if $x<0$ and $\mathrm{HD}(x)=0$ if $x\geq 0$.
We refer to the process of splitting a partial candidate codeword into two candidates by considering both the $0$ and $1$ hypotheses for the current bit as \emph{path forking}.
Only up to $L$ paths with the smallest PM values are retained after each~path~forking.

\subsection{Node-Based Decoding}\label{subsec:nodes}
To mitigate the huge latency when traversing the bottom of the decoding tree in SC and SCL decoding, special nodes identified by their distinct frozen bit patterns, as proposed in~\cite{Sarkis14Fast,Hanif17Fast,Condo18Generalized}, enable the direct computation of the PSUM vector $\bm{\beta}$ using tailored decoding algorithms.
To label whether a leaf bit in the fanout tree of a node $\mathcal{N}_{s,i}$ is frozen, we use a binary vector $\bm{d}_{s,i}$, where $\bm{d}_{s,i}[j]=0$ indicates a frozen bit and an information bit otherwise.
Following the constraints of~\cite{Ren22Sequence}, we consider five types of basic nodes whose vectors $\bm{d}_{s}$ are listed below.

\smallskip
\begin{tabular}{@{}p{6mm} @{}p{16mm} @{}l}
1) & \textbf{R0}   & $\bm{d}_s=(0,0,0,\dots,0,0)$ \tabularnewline
2) & \textbf{R1}   & $\bm{d}_s=(1,1,1,\dots,1,1)$ \tabularnewline
3) & \textbf{REP}  & $\bm{d}_s=(0,0,0,\dots,0,1)$ \tabularnewline
4) & \textbf{SPC}  & $\bm{d}_s=(0,1,1,\dots,1,1)$ \tabularnewline
5) & \textbf{TYPE-III}  & $\bm{d}_s=(\underbrace{0,0,1,\dots,1,1}_{2^s})$ \tabularnewline
\end{tabular}

Furthermore, the recent generalized SR node~\cite{Zheng21Threshold} includes most existing nodes as its special cases.
An SR node at stage $s$ consists of $s-r$ R0/REP nodes as its left descendants and a source node of any type as its last right-descendant located at lower stage $r$, where $r<s$.
Its pattern is shown in~\eqref{eq:SR_ds}, where each ``$\mathrm{X}$'' can be either a frozen or an information bit.
\begin{equation}\label{eq:SR_ds}
\bm{d}_s=(\underbrace{0, \dots, 0, \mathrm{X}}_{2^{s-1}}, \dots, \underbrace{0, \dots, 0, \mathrm{X}}_{2^{r}}, \underbrace{\mathrm{X}, \dots, \mathrm{X}}_{2^r})
\end{equation}

\begin{figure}[t]
\centering
\includegraphics[width=0.95\linewidth]{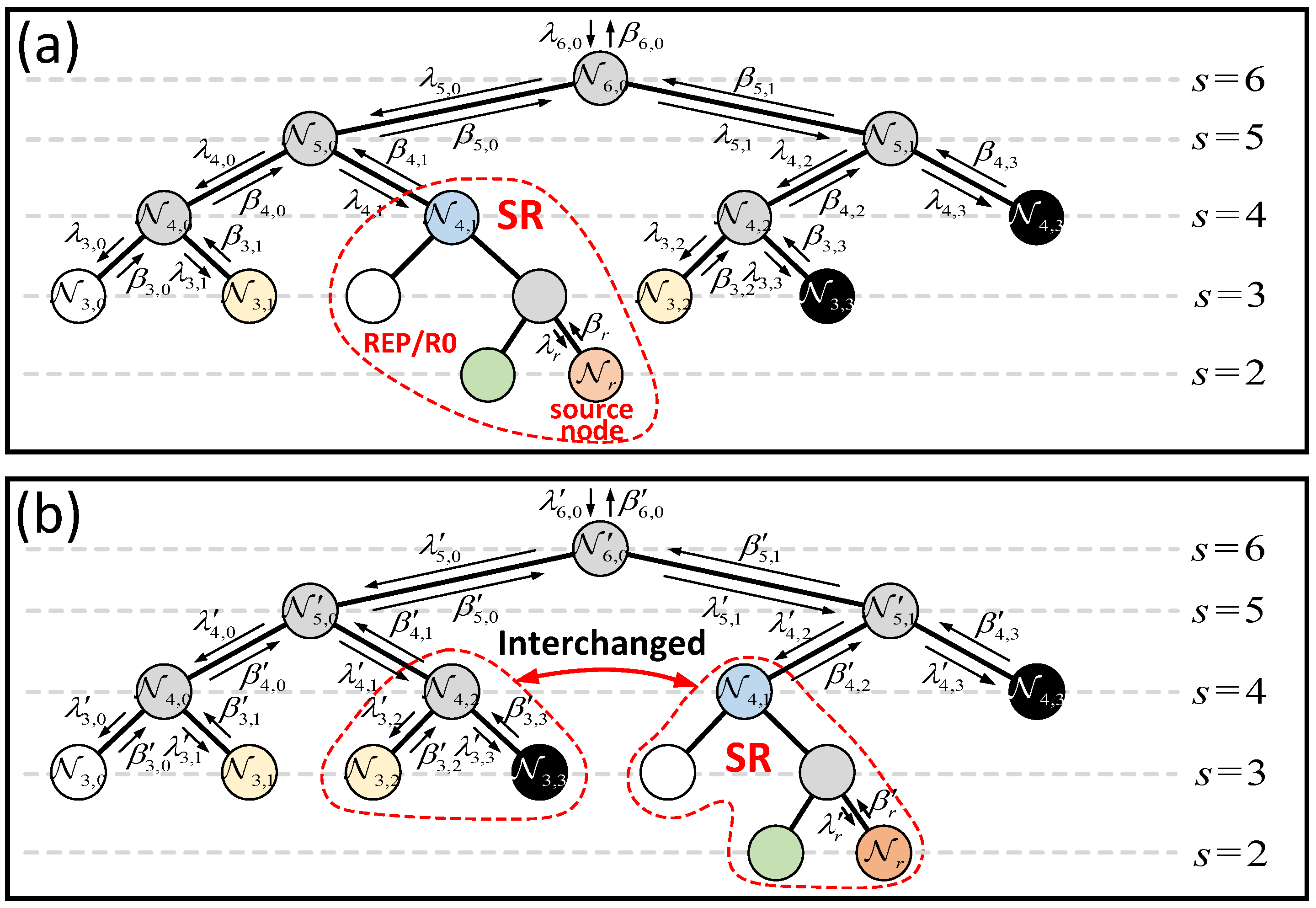}
\caption{(a) Decoding tree for $N=64$ polar codes with special nodes. (b) Decoding tree after factor graph permutation for $N=64$ polar codes. The node types are \tikzcircle[black, fill=\RZeroColor]{\tikzcirclewidth} for R0, \tikzcircle[black, fill=\myRepColor]{\tikzcirclewidth} for REP, \tikzcircle[black, fill=\ROneColor]{\tikzcirclewidth} for R1, \tikzcircle[black, fill=\mySpcColor]{\tikzcirclewidth} for SPC, \tikzcircle[black, fill=\TYPEIIIColor]{\tikzcirclewidth} for TYPE-III, \tikzcircle[black, fill=\mySRColor]{\tikzcirclewidth} for SR, and \tikzcircle[black, fill=shinygray]{\tikzcirclewidth} for a node of any rate. The superscript $^\prime$ indicates that the vectors $\bm{\lambda}$s and $\bm{\beta}$s are~permuted.}
\label{fig:TREE}
\end{figure}

As depicted in Fig.~\ref{fig:TREE}(a), $\mathcal{N}_{4,1}$ is an SR node that consists of two R0/REP nodes across stages $2$ to $3$ and a TYPE-III node as its source node $\mathcal{N}_r$ at stage $2$.
For brevity, we use SR(SNT) to describe an SR node, where SNT is the high-rate source node type restricted to R1, SPC, and TYPE-III in this work~\cite{Ren22Sequence}.
Assume that there exist $W$ R0/REP nodes in an SR node, $\mathbb{S}=\{\mathbb{S}_0,\mathbb{S}_1,\dots,\mathbb{S}_{2^{W}-1}\}$ denotes the set of all $|\mathbb{S}|$ possible repetition sequences for a given SR node.
It is noteworthy that thanks to its significant parallelism, the decoding of SR nodes provides low latency~\cite{Zheng21Threshold} and facilitates efficient hardware implementations of SCL decoding in 5G as~shown~in~\cite{Ren22Sequence}.
\begin{algorithm}[t]\footnotesize
%\SetKwInOut{Input}{Input}\SetKwInOut{Output}{Output}
\caption{\texttt{SCL Decoding with PFGs}()}
\label{alg:PSCL}
$\textbf{Input:}\;\bm{\lambda}_{n,0},\;\mathbb{A},\;\mathbb{P}$\\
$\bm{\hat{u}}\leftarrow\texttt{SCLdecoding}(\bm{\lambda}_{n,0}, \mathbb{A});$\\
\If{\upshape $\texttt{CRCdetection}(\bm{\hat{u}})==1$}
{
	\For{\upshape $i=1\;\textbf{to}\;|\mathbb{P}|-1$}
	{
		$[\bm{\lambda}_{n,0}^{\prime},\mathbb{A}^{\prime}]\leftarrow\texttt{permutation}(\bm{\lambda}_{n,0},\mathbb{A},\pi_i);$\\
		$\bm{\hat{u}^{\prime}}\leftarrow\texttt{SCLdecoding}(\bm{\lambda}_{n,0}^{\prime},\mathbb{A}^{\prime});$\\
		$\bm{\hat{u}}\leftarrow\texttt{permutationReverse}(\bm{\hat{u}^{\prime}},\;\pi_i);$\\
		\If{\upshape $\texttt{CRCdetection}(\bm{\hat{u}})==0$}
		{
			$\textbf{break}$\;
		}
	}
}
$\textbf{return}\;\bm{\hat{u}}$\;
\end{algorithm}

\subsection{Decoding with Permuted Factor Graphs}\label{sec:DecodePFG}
Decoding on PFGs can widen the codeword search space to further enhance the error-correcting capability of SCL decoding when the list size is constrained~\cite{Hashemi18Decoding}. 
Note that the codeword permutation proposed in~\cite{Doan18Decoding} reveals the one-to-one mapping between the permutation of factor graph stages and the shuffling of bit indices in a codeword. 
This enables the decoder to reuse its original architecture by merely shuffling the input LLRs to realize decoding on PFGs.
This idea has been implemented in the belief propagation list (BPL) decoder in~\cite{Ren24BPL}.
Herein, we outline the sequential processing of SCL decoding on multiple PFGs in Algorithm~\ref{alg:PSCL}. 
Given that the OFG $\OFG$ always yields the best average error-correcting performance~\cite{Doan18Decoding,Geiselhart20CRC}, the first decoding attempt is identical to the original SCL without any permutation. 
If decoding fails to satisfy the CRC, a codeword permutation is carried out to generate the permuted input LLRs $\bm{\lambda}_{n,0}^{\prime}$ as well as the corresponding information set $\mathbb{A}^{\prime}$ shown in Fig.~\ref{fig:TREE}(b), based on which the permuted SCL decoding is executed.
Let $\mathbb{P}$ denote the ensemble of $|\mathbb{P}|$ graph candidates, including $\OFG$ as its first element and $|\mathbb{P}|-1$ PFGs.
Once a decoding attempt passes the CRC detection, the decoded results are output.
Otherwise, the next attempt on a new PFG is scheduled until a valid codeword is generated or all the candidates in~$\mathbb{P}$~are~exhausted.
\begin{figure*}[t]
\centering
\includegraphics[width=0.725\linewidth]{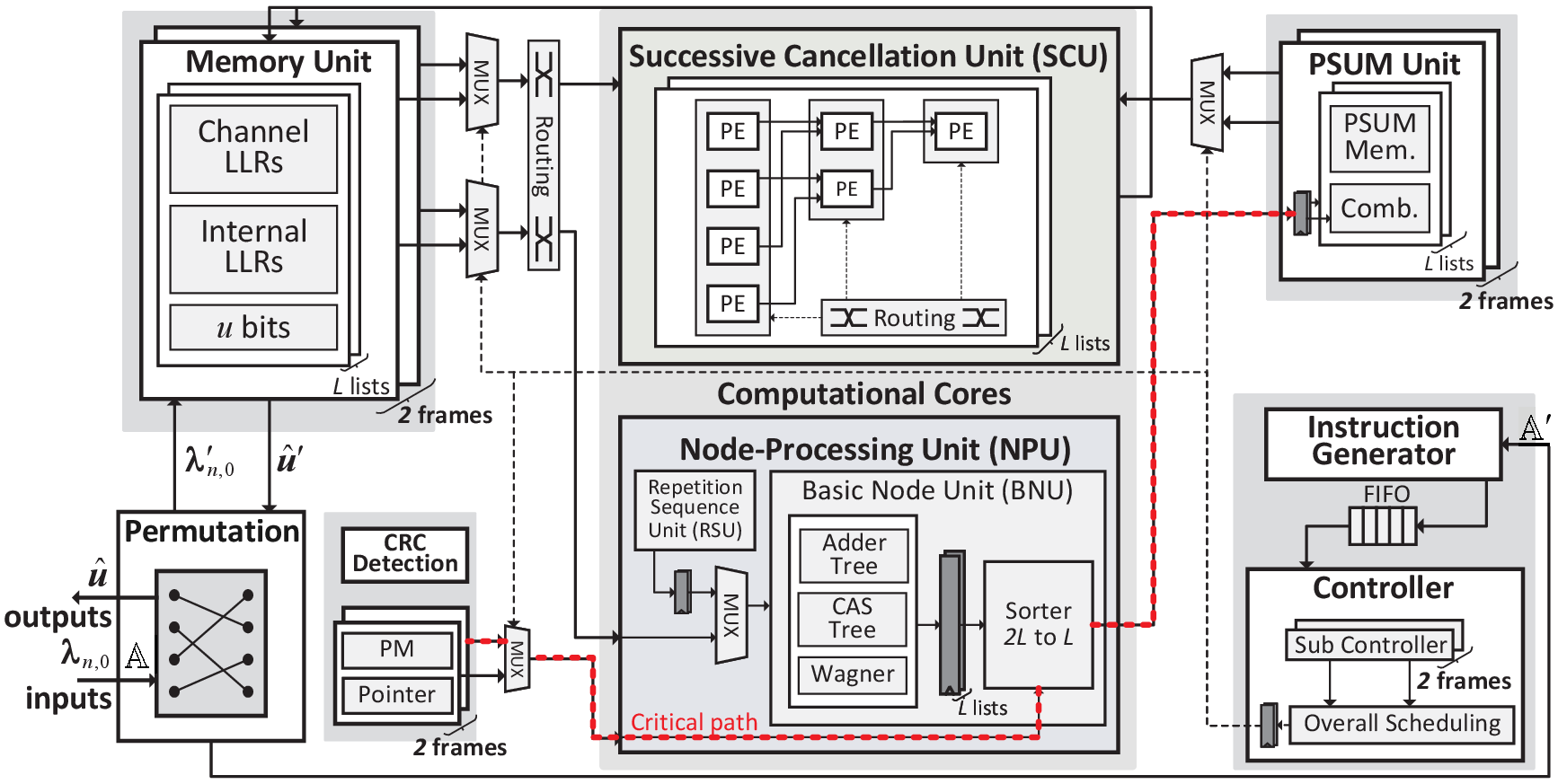}\\
\caption{Top-level overview of the proposed SCL decoder with interleaving architecture, where the red dotted lines show the critical path.}\label{fig:top}
\end{figure*}
\begin{figure}[t]
\centering
\includegraphics[width=1.0\linewidth]{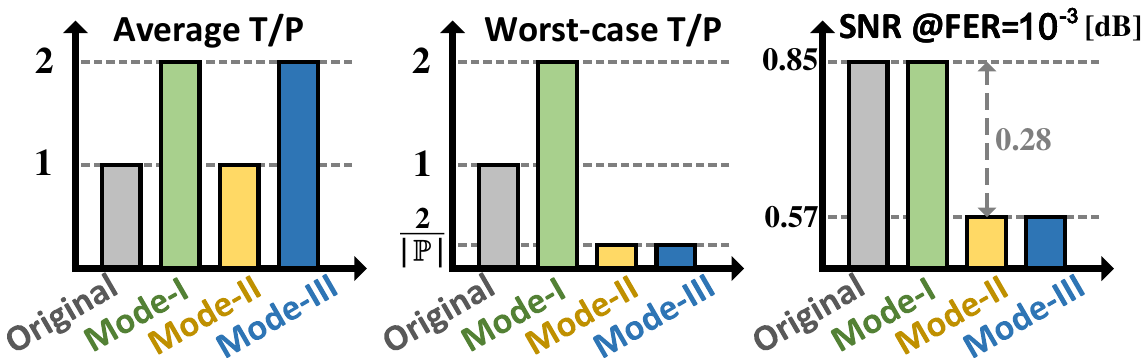}\\
\caption{Comparison between the baseline decoder~\cite{Ren22Sequence} and our frame-interleaving SCL decoder operating in various modes for DL-$(432,140)$ code using SCL-$8$ decoding with $|\mathbb{P}|=8$.}\label{fig:modeBar}
\end{figure}

\section{Hardware Architecture for Node-based SCL decoder with frame-interleaving}\label{sec:hardware}
Node-based decoding generally relies on two primary operations, which are mapped to two distinct hardware units in the decoder: a SC unit (SCU) for the internal LLR calculation and a node-processing unit (NPU) for fast node decoding.
However, due to the data dependency inherent in sequential processing, these two units must work in an alternating fashion.
Specifically, the NPU needs to wait until the SCU has computed the required LLR vector $\bm{\lambda}$. 
Similarly, the SCU requires the updated PSUM vector $\bm{\beta}$ from the NPU before proceeding. 
This sequential invocation always leaves one computational unit idle when the other one is running, limiting the hardware utilization and throughput. 
In~this section, we present a new node-based SCL decoder with a frame-interleaving architecture shown in Fig.~\ref{fig:top} to fill the resulting idle slots with another frame, thereby significantly enhancing hardware utilization.
Moreover, our decoder has three different working modes: Mode-I~Frame-Interleaving tailored to low latency, Mode-II~Graph-Interleaving tailored to high performance, and Mode-III~Hybrid-Interleaving (a mixture of the above two modes) tailored to high efficiency.
It is noteworthy that the internal controller can easily switch between~these three modes.
Implementation results shown in Fig.~\ref{fig:modeBar} demonstrate that our frame-interleaving SCL decoder operating across various modes has advantages in average throughput, worst-case throughput, and reliability compared to the baseline decoder~\cite{Ren22Sequence}.
This frame-interleaving architecture is especially useful for the blind decoding and detection required for 5G where the received data must be decoded multiple times speculating on the transmit/encoding parameters~\cite{condo2018design}.
It is further useful for high-throughput data channels where after a short initial latency, two frames can be decoded in almost the same time as a single frame with a non-interleaved decoder.

\subsection{Top-Level Hardware Architecture}\label{subsec:top}
In Fig.~\ref{fig:top}, we present a top-level illustration of our SCL decoder with interleaving architecture.
This architecture comprises four main types of modules: computational cores, memories, a permutation generator~\cite{Ren24BPL}, and a controller with an online instruction generator.
Since we intend to process two independent frames concurrently in an interleaved fashion, storage (memories) in the decoder must be duplicated, including channel LLRs, internal LLRs, output bits, PSUMs, pointers, and PMs.
The computational cores (i.e., the SCU and the NPU) read channel or internal LLRs from the memory unit to calculate the vectors $\bm{\lambda}$s and $\bm{\beta}$s, respectively. 
Specifically, the SCU is implemented by multiple stages of processing elements (PEs)~\cite{liu20185} to calculate the internal LLRs $\bm{\lambda}$ (referring to~\eqref{eq:llr_fg_func}), processing several stages of the decoding tree in a single cycle.
This technique can significantly reduce the internal LLR memory overhead since intermediate LLRs are re-calculated combinatorially without being stored~\cite{liu20185}.
Note that the SCU operation continues until a special node is encountered, and the resulting LLRs $\bm{\lambda}$ are then passed to the NPU for node-based operations.
Meanwhile, to support list decoding for SR nodes~\cite{Ren22Sequence}, the NPU instantiates two core units, a repetition sequence unit (RSU) for the low-rate part of SR nodes (enumerates all possibilities and then performs $|\mathbb{S}|L\rightarrow L$ sorting) and a basic node unit (BNU) for the high-rate source node (sequentially runs multiple $2L\rightarrow L$ path forking).
The BNU consists of an adder tree, a compare-and-select (CAS) tree, a Wagner decoder for parity check, and a $2L$-to-$L$~sorter for path forking~\cite{Bala15LLR}.
Note that the processing of a single node can take multiple clock cycles~(CCs), which is discussed further in Section~\ref{sec:strategy}.
\begin{figure}[t]
\centering
\includegraphics[width=1.0\linewidth]{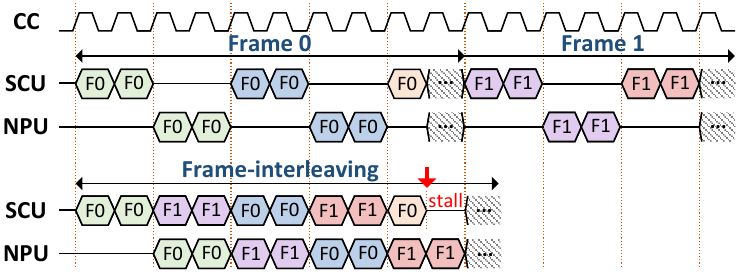}\\
\caption{Decoding schedules of two frames using conventional single-frame architecture and our interleaving architecture, where the $i$-th frame is denoted as $\mathsf{F}i$. Operations on different nodes are marked by distinct colors for clarity.}\label{fig:frameInterleaving}
\end{figure}
\begin{figure}[t]
\centering
\includegraphics[width=1.0\linewidth]{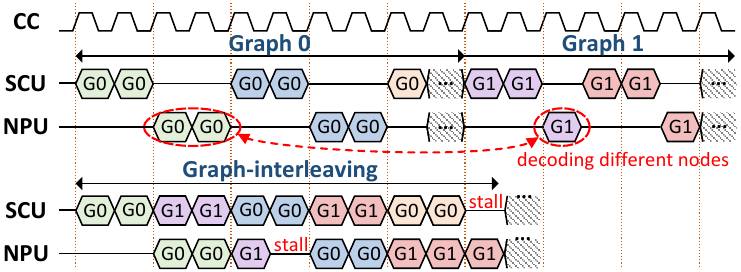}\\
\caption{Decoding schedules of two graphs on the same frame using conventional single-frame architecture and our interleaving architecture, where the $i$-th graph is denoted as $\mathsf{G}i$.}\label{fig:graphInterleaving}
\end{figure}

Our frame-interleaving architecture time-shares the two computational units (the SCU and the NPU) between two independent frames.
We denote them as $\FrameX$ and $\FrameY$ for brevity.
Given that the SCU and the NPU typically occupy more than half of the area~\cite{Ren22Sequence}, time-sharing these computational units offers an obvious area advantage for our interleaving architecture, compared to a simple instantiation of two parallel decoders (multiple cores).
While the SCU is calculating LLRs for $\FrameX$, the NPU performs node list decoding for $\FrameY$ and vice versa.
The controller generates the control logic to determine whether to read the data for either $\FrameX$ or $\FrameY$, which contains two sub-controllers and a scheduling unit.
Two sub-controllers orchestrate the decoding process of two frames independently, and the scheduling unit is responsible for the hand-shaking among two sub-controllers and the datapath components of the decoder.
An online instruction generator avoids the need for a memory to store all offline-generated instructions (e.g., in the non-interleaving baseline SCL decoder~\cite{Ren22Sequence}).
This generator can identify the node patterns (including SR nodes) for any given information set and delivers a list of instructions on the fly, greatly enhancing the flexibility of our SCL decoder and reducing the area overhead of the instruction memory.
We describe it in~detail~in~Section~\ref{sec:instruction}.

\subsection{Timing and Operation Schedules for Different Modes}\label{subsec:modes}
In this section, we give example timing diagrams to clearly illustrate the scheduling of our three working modes.\footnote{Note that the PSUM unit~(PSU) consistently consumes one cycle to execute~\eqref{eq:psum_func} after the NPU processing~\cite{Ren22Sequence}. For brevity, this cycle is conceptually integrated into the SCU processing time for the subsequent node. This allows us to focus on the SCU and the NPU as the primary time-consuming~units.}

\subsubsection{Mode-I Frame-Interleaving}\label{subsubsec:frame}
In this mode, we interleave two frames to fill the unused slots in the SCU and the NPU while waiting for the other unit to complete its operation.
As illustrated in Fig.~\ref{fig:frameInterleaving}, the decoding process of~$\FrameY$ is moved forward and integrated into the idle cycles caused by~$\FrameX$.
Ideally, both frames proceed in a fully pipelined manner without any idle cycles of the SCU and the NPU. 
Yet, in practice, some inevitable stalls occur due to the mismatch in the number of processing cycles of the SCU and the NPU.
For length-$432$ DL polar codes, the average and worst-case number of residual stalls in Mode-I are $21.3$ and $29$ cycles, respectively. 
For length-$1024$ UL polar codes, these numbers increase to $58.3$ and $78$ cycles, respectively.
For example, as shown by the red arrow in Fig.~\ref{fig:frameInterleaving}, $\FrameX$ finishes the $\bm{\lambda}$ calculation and then asserts the request signal to the NPU.
The overall scheduling unit (shown in Fig.~\ref{fig:top}) performs hand-shaking operations with the NPU, but discovers that it is busy processing~$\FrameY$. 
Hence, $\FrameX$ has to wait until the NPU is idle, thus leading to a stall.
Nevertheless, the resulting stalls are much fewer than those in the traditional single-frame decoder, thereby resulting in better hardware utilization as well as higher (both average and worst-case) throughput.

\subsubsection{Mode-II Graph-Interleaving}\label{subsubsec:graph}
Our decoder can employ graph permutations on the same codeword to widen the decoding search space.
As mentioned in Section~\ref{sec:DecodePFG}, decoding on PFGs can further enhance the error-correcting capability of the SCL decoder with a constrained list size.
This is similar to AED~\cite{geiselhart2021automorphism}, but we only utilize factor graph permutation~\cite{Elke18Belief,Ren2020TVT}, a subset of the GA group~\cite{geiselhart20236g} that can be implemented efficiently in hardware~\cite{Ren24BPL}.
To balance hardware efficiency and throughput, rather than instantiating multiple sub-decoders like~\cite{johannsen2023successive}, we implement a serial decoding scheme that shuffles the input LLRs and reuses a single decoder~\cite{Doan18Decoding,Ren24BPL} with frame-interleaving, as outlined in Algorithm~\ref{alg:PSCL}.
While the serial scheme suffers from a longer worst-case decoding latency (increasing proportionally with $|\mathbb{P}|$), our decoder can deliver the same error rate as the parallel scheme~\cite{johannsen2023successive} at a minor cost of additional area for permutation~generation.

For each graph, we adopt a flexible permutation generator~\cite{Ren24BPL} to shuffle the information set~$\mathbb{A}$ and the input LLRs of each frame.
Note that the shuffling of the input LLRs alters the bit indices within the factor graph and consequently greatly affects the node structure of the decoding tree~\cite{Doan18Decoding,Hashemi18Decoding}.
Unfortunately, this alternation generally complicates the direct application of node techniques to graph ensemble decoding.
However, inspired by~\cite{geiselhart2021automorphism}, we can perform a partially ordered codeword permutation to retain the underlying node structures.
Specifically, when permuting the decoding tree (factor graph), we maintain the bottom stages fixed but only permute the upper stages of the decoding tree.
This approach can effectively combine the advantages of both graph ensemble decoding and node-based fast decoding.

For instance, Fig.~\ref{fig:TREE}(b) illustrates an example of a permuted decoding tree, where the lower four stages are fixed.
Bit indices are exchanged in groups of length $2^4=16$ while the internal order within each group is not disrupted.
Compared to the original decoding tree, two nodes at stage $s = 4$ swap positions accordingly in the permuted tree, marked as $\mathcal{N}_{4,1}\leftrightarrow \mathcal{N}_{4,2}$.
It is noteworthy that if we consider graph ensemble decoding as decoding multiple frames containing the same set of special nodes, but shuffled in their order, a ``graph'' can be naturally regarded as a ``frame'' for our frame-interleaving decoder.
In Fig.~\ref{fig:graphInterleaving}, the proposed architecture can thus process two graphs for the same frame simultaneously to notably reduce the worst-case latency in the graph-interleaving~mode.
\begin{figure}[t]
\centering
\includegraphics[width=1.0\linewidth]{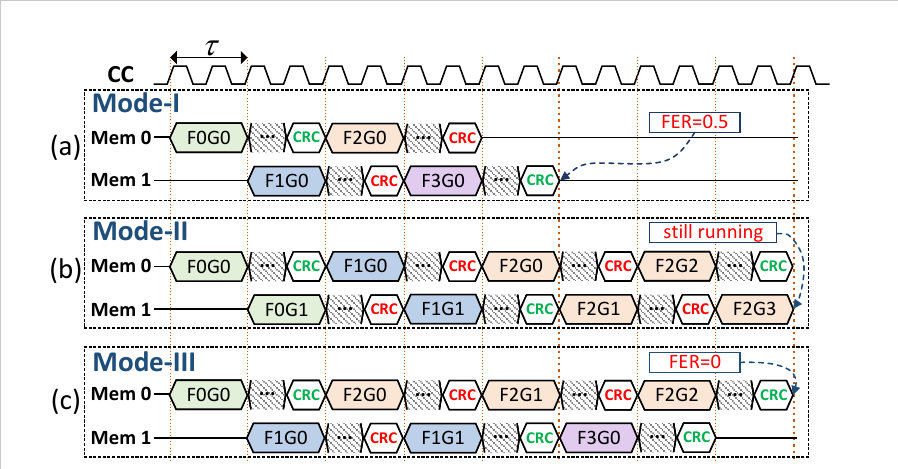}\\
\caption{Decoding schedules of three working modes from the perspective of the memory occupation, where the $i$-th frame on the $j$-th graph is denoted as $\mathsf{F}i\mathsf{G}j$. Herein, operations on different frames are marked by distinct colors. The green $\mathsf{CRC}$ represents the CRC check succeeds otherwise labeled as~red.}\label{fig:hybridInterleaving}
\end{figure}

\subsubsection{Mode-III Hybrid-Interleaving}\label{subsubsec:hybrid}
Mode-III merges the benefits of both Mode-I and Mode-II.
Fig.~\ref{fig:hybridInterleaving} illustrates the memory occupation for these modes.\footnote{In our decoder, one cycle is necessary for the CRC detection. Besides, the shuffling of input LLRs will not introduce additional delays as the permutations shown in~Fig.~\ref{fig:top} and the decoder operate in a fully pipelined fashion. Specifically, while the decoder processes SCL decoding on~$\pi_{i}$, the permutation generator is already preparing the next permutation~$\pi_{i+1}$~\cite{Ren24BPL}.}
Let $\MemX$ and $\MemY$ denote two separate memory instances in the proposed architecture.
As mentioned in Section~\ref{subsubsec:frame}, Mode-I focuses on decoder throughput improvement rather than better error-correcting performance.
It thus only continuously interleaves two distinct frames on the OFG without using any other PFGs.
Notably, $\mytextsf{F1G0}$ must wait for several cycles until $\mytextsf{F0G0}$ completes the LLR calculation for the first node and then releases the SCU.
For the sake of generality, we assume this stall lasts $\tau$ cycles.
Conversely, Mode-II prioritizes reducing the error rate through graph ensemble decoding by interleaving as many graphs as possible.
Nevertheless, as most erroneous frames in the medium to high SNR region are corrected by the OFG, there are many parallel redundant attempts in Mode-II.
For instance, our decoder launches $\mytextsf{F0G0}$ and $\mytextsf{F0G1}$ simultaneously in Fig.~\ref{fig:hybridInterleaving}, yet, as $\mytextsf{F0G0}$ passes the CRC detection while $\mytextsf{F0G1}$ fails, the decoding attempt on $\mytextsf{F0G1}$ was effectively useless.
Therefore, while Mode-II delivers superior reliability than Mode-I by allowing for up to $|\mathbb{P}|$ additional decoding attempts, its \emph{average throughput} is slightly less than half of Mode-I and approximately identical to the baseline decoder.

To address this issue, we modify the decoding schedule and propose Mode-III (hybrid-interleaving) as a mixture of the above two modes.
Two features of Mode-III are summarized as:
\circled[0.8]{1}~decoding two distinct frames at any one time; \circled[0.8]{2}~switching to the next graph only if the current decoding attempt fails.
We give priority to executing frame-interleaving and allow for additional decoding attempts on more graphs only when needed.
As illustrated in Fig.~\ref{fig:hybridInterleaving}(c), similar to Mode-I, we first decode $\mytextsf{F0G0}$ and $\mytextsf{F1G0}$.
Once $\mytextsf{F0G0}$ succeeds, decoding on $\mytextsf{F2G0}$ is launched.
Only when $\mytextsf{F1G0}$ fails, an additional attempt for the current frame on $\mytextsf{G1}$ is required. 
Note that in this case, the order of sequential decoding outputs cannot be guaranteed (e.g., $\mytextsf{F3}$ outputs before $\mytextsf{F2}$ in Fig.~\ref{fig:hybridInterleaving}(c)) due to the fact that some frames may need time-consuming serial decoding attempts on more graphs to derive a valid result.

Fig.~\ref{fig:modeBar} provides a comparison between our proposed decoder with three interleaving modes and the baseline decoder~\cite{Ren22Sequence} in terms of the average throughput, the worst-case throughput, and the SNR at the target of $10^{-3}$.
In ideal scenarios without any stalls (i.e., no mismatch between the SCU and the NPU), Mode-I and Mode-III enjoy a $2\times$ average throughput compared to~\cite{Ren22Sequence}.
However, Mode-II has no improvement in this aspect due to the unnecessary decoding attempts mentioned above.
Besides, serial graph ensemble decoding also impairs the worst-case latency, thus leading to a $|\mathbb{P}|\times$ latency increase.
Nevertheless, thanks to the proposed interleaving architecture, Mode-II and Mode-III halve the worst-case latency by decoding two graphs simultaneously, resulting in a $\frac{2}{|\mathbb{P}|}\times$ worst-case throughput compared to the baseline decoder~\cite{Ren22Sequence}.
Notably, a significant error-correcting performance gain is achieved by Mode-II and Mode-III.
For DL-$(432, 140)$ code, the SNR at the target of $10^{-3}$ is lowered by $0.28$ dB using SCL-$8$ decoding with $|\mathbb{P}|=8$.
In summary, Mode-III showcases the advantages in both average throughput and reliability, which renders it a preferable operating pattern based on the proposed~architecture.

\section{Methods for Eliminating Remaining Stalls in the Interleaving Schedule}\label{sec:strategy}
Ideally, the proposed architecture time-shares two computational units and fills all the idle slots with another frame (or graph) to enhance hardware utilization by $2\times$.
However, as mentioned in Section~\ref{subsec:modes}, the unavoidable stalls from the mismatch in the processing cycles of the SCU and the NPU remain, which is a significant challenge.
The dominating mismatch pattern arises from the necessity for the SCU to idle while waiting for the NPU to decode high-rate nodes. 
This NPU process can be time-consuming as it sequentially executes multiple $2L\rightarrow L$ path forking steps.
In this context, we present two strategies ($\SX$ and $\SY$) to eliminate these remaining stalls.
Note that both $\SX$ and $\SY$ only require changes of control signals, which is more convenient than~\cite{Kam23low} to realize overlapped operations and avoid altering the architectures of computational units.
\begin{figure}[t]
	\centering
	\includegraphics[width=1.0\linewidth]{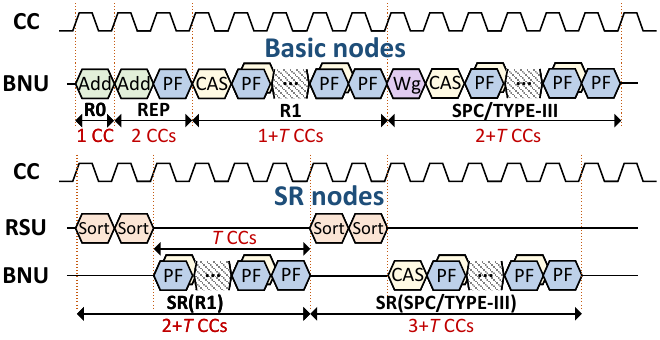}\\
	\caption{Cycle-level analysis of node-based decoding for basic and SR nodes~\cite{Ren22Sequence}. The terms $\mathsf{Add}$, $\mathsf{PF}$, and $\mathsf{Wg}$ represent the addition, path forking, and Wagner decoding, respectively.}\label{fig:nodeCC}
\end{figure}
\begin{figure}[t]
\centering
\includegraphics[width=1.0\linewidth]{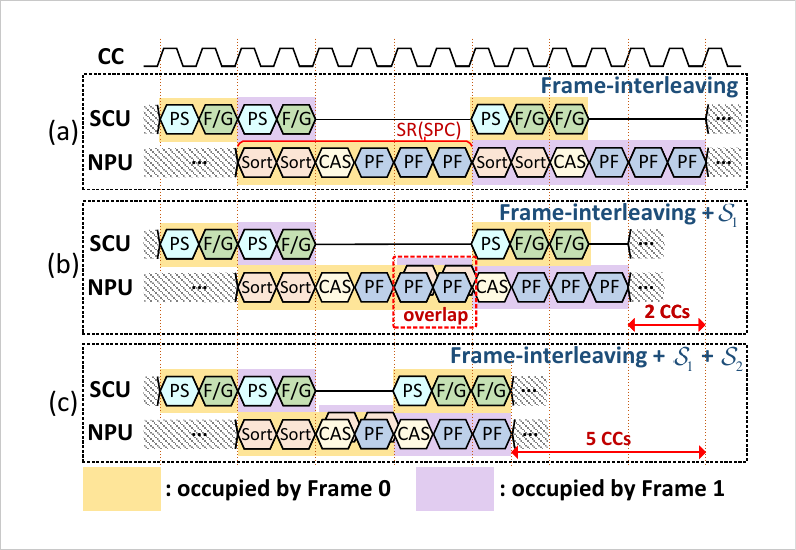}\\
\caption{Decoding schedule of our frame-interleaving decoder adopting two stall-eliminating strategies. The terms $\mathsf{PS}$ and $\mathsf{F/G}$ represent the PSUM combine and the input LLR calculation of special nodes, respectively.}\label{fig:strategy}
\end{figure}

\begin{figure}[t]
\centering
\begin{minipage}[t]{\columnwidth}
	\centering
	\includegraphics[width=\columnwidth]{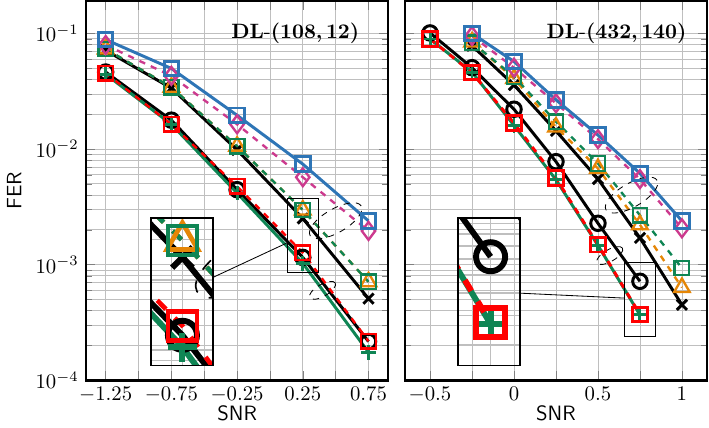}
	\label{fig:UL_FER}
\end{minipage}
\vspace*{-0.25cm}

\begin{minipage}[t]{\columnwidth}
	\centering
	\includegraphics[width=\columnwidth]{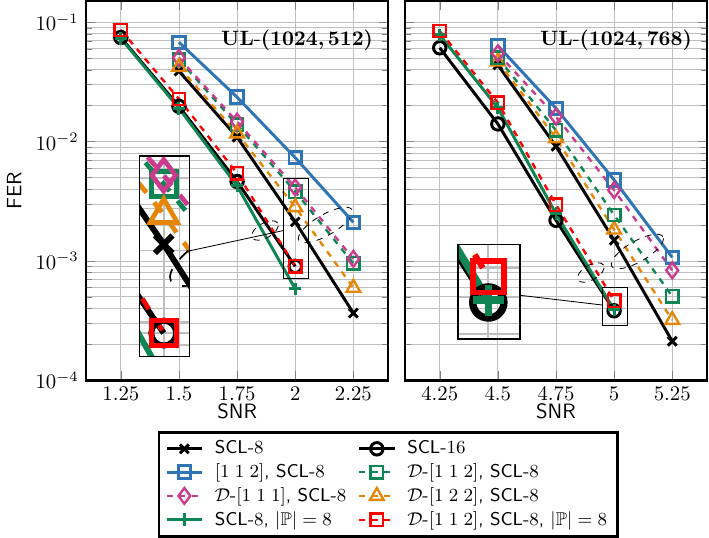}
	\label{fig:DL_FER}
\end{minipage}
\caption{Quantized FER performance of SCL decoding with different configurations.}
\label{fig:FER}
\end{figure}

\subsection{Cycle-Level Analysis of the Node-Based Processing}\label{subsec:cycleAnalysis}
In this section, we first recall the detailed schedule of the NPU running fast list decoding for distinct nodes, as shown in Fig.~\ref{fig:nodeCC}.
For R0 and REP nodes, an adder tree takes one CC to compute the PM increment, while the path forking of the single information bit costs an additional cycle in REP nodes.
For high-rate nodes, sequential path forking is executed in ascending order of the reliability of the information bits ($|\bm{\lambda}_s[i]|,i\in[0,2^{s}-1]$).
To balance decoding latency and error rate in this decoder, we empirically assign a constraint $T$ to the number of path forks\cite{Hashemi17Fast}.\footnote{The accurate number of path forks equals to $\min(T,K_s)$, where $K_s$ is the number of unfrozen bits in the node.}
Numerical results demonstrate that a reasonable value for $T$ can significantly reduce decoding latency with negligible error-rate performance degradation~\cite{Ren22Sequence}.
Therefore, an R1 node takes up to $T+1$ CCs, and an SPC/TYPE-III node needs  an extra CC for Wagner decoding to first check for the global/even parity-check constraint, thus leading to a latency of $T+2$ CCs.
In terms of the SR nodes, the RSU takes a fixed two CCs to perform the $|\mathbb{S}|L\rightarrow L$ large-size sorting~\cite{Ren22Sequence} (as mentioned in Section~\ref{subsec:nodes}).
It is noteworthy that the first CAS cycle for the SR(R1) node and the first Wagner decoding for the SR(SPC/TYPE-III) are merged in the RSU to effectively reduce decoding cycles~\cite{Ren22Sequence}.\footnote{We use the term SR(R1/SPC/TYPE-III) to denote the SR node with a source node of R1/SPC/TYPE-III.}

\subsection{$\SX$: Dynamic Overlapped Processing of SR Nodes}\label{subsec:S1}
In this section, we introduce the $\SX$ strategy, a dynamic overlapped processing for efficiently decoding time-consuming SR nodes by exploiting the independence of the RSU from the BNU.
Fig.~\ref{fig:strategy}(a) illustrates a partial decoding schedule running over an SR(SPC) node for $\FrameX$ and $\FrameY$.
According to the alternating fashion, $\FrameY$ needs to wait for the idle NPU until $\FrameX$ completes the SR(SPC) node processing, leading to a delay of $4$ CCs.
However, as mentioned in Section~\ref{subsec:top}, the NPU consists of two units that are separated by a pipeline stage: the RSU and the BNU.
By pre-loading $\FrameY$ to the RSU to decode the low-rate part of the SR(SPC) node, concurrent with path forking of $\FrameX$ in the BSU, two stalls CCs of $\FrameY$ can thus be saved.
Note that the $\SX$ strategy is contingent upon two conditions to execute: \circled[0.8]{1}~the processing node must be an SR node, \circled[0.8]{2}~$\SX$ must not bring new delays.
Namely, ensuring that the current NPU processing delays are at least $2$~CCs so that the RSU can overlap with these delays.
Consequently, the $\SX$ strategy effectively reduces stalls without any error-rate performance degradation, decreasing the overall latency by $5\%$ for UL-$(1024,512)$ polar code compared to the original frame-interleaving design~\cite{Zhang2024ISCAS}.
\begin{figure}[t]
	\centering
	\includegraphics[width=0.9\linewidth]{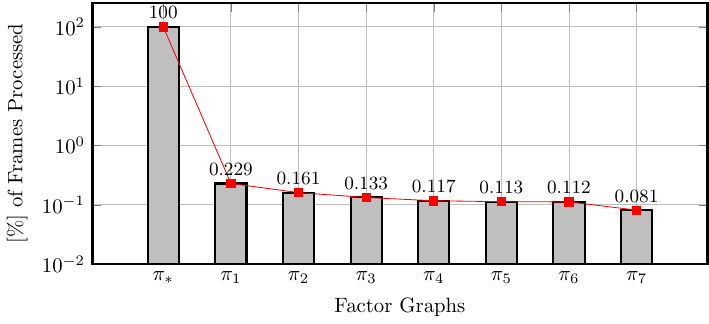}
	\caption{Percentage histogram of frames processed by each factor graph using SCL-$8$ decoding with $\mathcal{D}$-$[1,1,2]$, $|\mathbb{P}|=8$ for UL-$(1024,512)$.}\label{fig:histogramFG}
\end{figure}

\subsection{$\SY$: Dynamic Path Forking for High-Rate Nodes}\label{subsec:S2}
To further eliminate residual stalls in our frame-interleaving SCL decoder, we propose an additional optimization called the $\SY$ strategy.
This method dynamically reduces stalls resulting from the path forking of high-rate nodes while maximizing the number of path forks during periods when the SCU is busy.
This approach carefully balances latency reduction with minimal impact on performance.
As shown in the example schedule of Fig.~\ref{fig:strategy}(c), $\FrameX$ reduces its path forks by two, and $\FrameY$ by one, for the SR(SPC) node, collectively achieving a $3$-CC reduction compared to the $\SX$ strategy.
Note that while the $\SY$ strategy offers a notable reduction in latency, it does not allow for an infinite reduction in path forking.
Herein, we introduce a lower bound on path forking for special nodes to guarantee minimum numbers of path forks in the $\SY$ strategy.
This constraint ensures that our decoder maintains an acceptable level of performance degradation, which is discussed below.
\begin{figure}[t]
	\centering
	\begin{minipage}[t]{\columnwidth}
		\centering
		\includegraphics[width=\columnwidth]{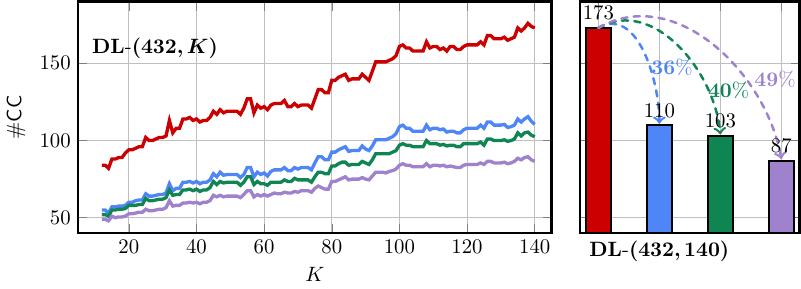}
		\label{fig:UL_lat}
	\end{minipage}
	\vspace*{-0.25cm}
	
	\begin{minipage}[t]{\columnwidth}
		\centering
		\includegraphics[width=\columnwidth]{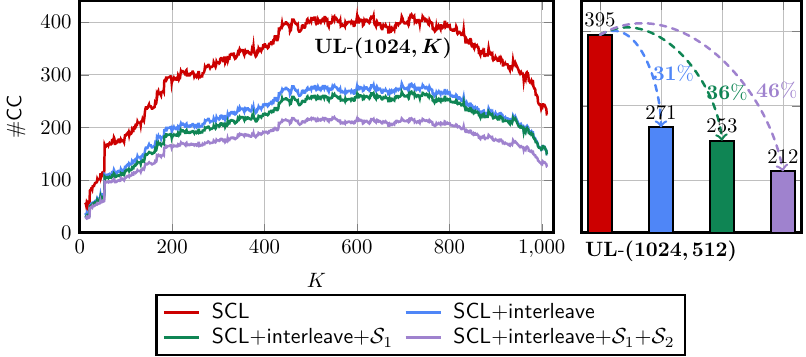}
		\label{fig:DL_lat}
	\end{minipage}
	\caption{Average latency analysis of different SCL decoders for DL-$(432,K)$ and UL-$(1024,K)$ codes with $L=8$, where the baseline SCL refers to~\cite{Ren22Sequence}.}
	\label{fig:avgLatency}
\end{figure}

We define a vector $\mathcal{T}=[T_{\text{R1}},T_{\text{SPC}},T_{\text{TYPE-III}}]$ to denote the empirical number of path forks for R1, SPC, and TYPE-III nodes, including SR nodes using these as source nodes.
It is noteworthy that~\cite{Ren22Sequence} demonstrates that a default setting of $\mathcal{T}=[2,3,3]$ is a very good latency-performance trade-off for baseline SCL-$8$ decoding.
Then, Fig.~\ref{fig:FER} illustrates the quantized FER performance for SCL decoding across different $\mathcal{T}$ for DL and UL codes, where $\mathcal{D}\text{-}\mathcal{T}$ denotes the $\SY$ strategy with $\mathcal{T}$ as its lower bound.
In this work, we utilize a sign-magnitude quantization with the $6$-bit LLRs and $7$-bit PMs, which has a negligible FER performance loss compared to floating-point.
Fig.~\ref{fig:FER} reveals that the error-correcting performance gap widens as the lower bound decreases (i.e., reducing more path forking), relative to baseline SCL-$8$ decoding.
To obtain an optimal latency-performance balance, the $\mathcal{D}\text{-}[1,1,2]$ configuration emerges as the preferred choice, with a performance loss of less than $0.15$~dB across all simulations. 
This dynamic configuration retains more path forks than the static $[1,1,2]$ setup, offering a notable performance advantage with the same latency.
Specifically, for DL-$(432, 140)$ codes, SCL-$8$ decoding using $\mathcal{D}\text{-}[1,1,2]$ surpasses the static $[1,1,2]$ configuration by approximately $0.25$~dB.
This improvement is attributed to $35\%$ more path forks by the $\SY$ strategy without causing additional stalls. 
For graph ensemble decoding with $8$ graphs, $\mathcal{D}\text{-}[1,1,2]$ incurs a minor performance loss, but still maintains comparable error-correcting capacity to SCL-$16$~decoding in Fig.~\ref{fig:FER}.
Moreover, Fig.~\ref{fig:histogramFG} illustrates a percentage histogram of frames processed by each factor graph using SCL-$8$ decoding with $\mathcal{D}$-$[1,1,2]$, $|\mathbb{P}|=8$ for UL-$(1024,512)$ when the target FER is $10^{-3}$.
After the OFG, the usage of $\pi_{1}$ is only $0.229\%$.

\subsection{Decoding Latency Analysis}
\begin{figure}[t]
\centering
\begin{minipage}[t]{\columnwidth}
	\centering
	\includegraphics[width=0.95\columnwidth]{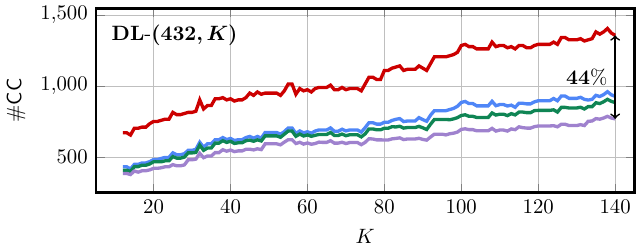}
	\label{fig:UL_wclat}
\end{minipage}
\vspace*{-0.4cm}

\begin{minipage}[t]{\columnwidth}
	\centering
	\hspace*{0.11cm}
	\includegraphics[width=0.95\columnwidth]{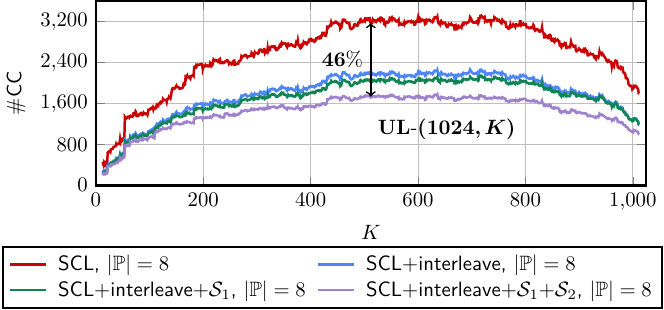}
	\label{fig:DL_wclat}
\end{minipage}
\caption{Worst-case latency analysis of SCL-$8$ decoders for DL-$(432,K)$ and UL-$(1024,K)$ codes with $8$ graphs per frame.}
\label{fig:wcLatency}
\end{figure}

To assess the latency reductions achieved by our proposed methods across various scenarios, Fig.~\ref{fig:avgLatency} plots the average decoding cycle count per frame for all message lengths $K$ in DL-$(432,K)$ and UL-$(1024,K)$ with different SCL decoders.
Following the approach in Section~\ref{subsec:S2}, we adopt $\mathcal{D}$-$[1,1,2]$ as the $\SY$ configuration in the following latency analysis.
When decoding the DL-$(432,140)$ code, the frame-interleaving architecture alone can achieve a $36\%$ latency reduction over the baseline decoder~\cite{Ren22Sequence}.
Moreover, the $\SX$ and $\SY$ strategies can further deliver additional latency decreases of $4\%$ and $9\%$, respectively, nearly cutting the decoding time into half (around $50\%$) to only $87$~CCs.
Similarly, for the UL-$(1024,512)$ code, there is a cumulative latency decrease of $46\%$, resulting in only $212$~CCs per frame on average.
These results underscore the positive impact of our optimization strategies across various code rates and lengths, which nearly achieve the ideal target of the proposed architecture (i.e., time-sharing two computational units, filling all idle slots).
Fig.~\ref{fig:wcLatency} illustrates the worst-case latency for graph ensemble decoding.
When handling $8$ graphs, the baseline SCL-$8$ decoder typically suffers from an $8$-fold increase in latency (indicated by the red~line).
However, with the proposed decoder, we can reduce the worst-case latency by $44\%$ and $46\%$ for DL-$(432,140)$ and UL-$(1024,512)$ codes,~respectively.

\section{Online Instruction Generator}\label{sec:instruction}
Node-based polar decoders rely on identifying the types and lengths of special nodes. 
This necessitates complex control logic.
To bypass this issue, most decoders perform this identification offline and then store the resulting instructions in dedicated memories. 
The decoder then proceeds by reading these instructions to coordinate each component. 
However, this method incurs significant area overhead for storing instructions and makes it almost impossible to rapidly change the rate as required by 5G~\cite{Ren22Sequence,ren2024generalized}.
Note that these instructions~are~closely tied to $N$ and $K$, and any changes to these parameters require a new list of instructions.
As the range of message lengths is from $12$ to $1706$ in 5G~NR polar codes, this offline generation can not be precomputed for all relevant code configurations.
Moreover, given the complexity of our frame-interleaving decoder, we also support three distinct working modes, graph ensemble decoding, generalized SR nodes, and multiple dynamic processing strategies to optimize the performance and latency. 
These factors strongly motivate the design of an online instruction generator to enable the required flexibility.

In this section, we design an online instruction generator tailored to work synchronously with the decoder to identify basic nodes and generalized SR nodes and generate the corresponding instructions on-the-fly.
Instead of storing a bit-channel reliability vector as in~\cite{hashemi2019rate}, our generator operates directly with the binary sequence defining the information bits $\mathbb{A}$ (or $\mathbb{A}^{\prime}$ for PFGs).\footnote{For brevity, we still represent the binary sequence vector of the information set as $\mathbb{A}$, where $\mathbb{A}[i]$ indicates a frozen bit if $0$ and an information bit as $1$.}
It features a sliding window of maximum node size $N_{s_{\max}}$, moving along the sequence $\mathbb{A}$.
This window contains a sub-sequence $\mathbb{A}_{\text{sub}}$, enabling the identification of special nodes at varying lengths within $\mathbb{A}_{\text{sub}}$, to progress until the entire sequence $\mathbb{A}$ is analyzed.

\subsection{Node Identification for Basic Nodes}\label{subsec:id_basic}
Given the frozen bit patterns of the five basic nodes described in Section~\ref{subsec:nodes}, we group the basic nodes as \{R0, REP\} (special cases of generalized REP nodes~\cite{Condo18Generalized}) and \{R1, SPC, TYPE-III\} (special cases of generalized parity-check nodes~\cite{Condo18Generalized}), as each group shares similar patterns for their node recognition independent of their length.
Hence, we employ a unified logic for each group in pursuit of area saving.
We define a symbol $\mathcal{M}^k_{\text{X}}$ to track whether a length-$2^{k}$ special node ``X'' is detected or not in the sub-sequence $\mathbb{A}_{\text{sub}}$ corresponding to the sliding window, where ``X'' means the node type and $\log N_{s_{\min}}\leq k< \log N_{s_{\max}}$.
For the \{R0, REP\} group, the control logic can be mathematically formulated as
\begin{equation}\label{eq:logic_r0_rep}
	\begin{small}
		\begin{aligned}
			&\mathcal{M}^{k+1}_{\text{R0}}\!\!\!\!\!\!&&=\left\{
			\begin{aligned}
				&1, \text{ if }\mathcal{M}^{k}_{\text{R0}}\!=\!1\text{ and }\mathbb{A}_{\text{sub}}[2^{k+1}\!-\!1]\!=\!0\text{ and }\xi\,, \\
				&0, \text{ otherwise} \,,
			\end{aligned} \right.\\
			&\mathcal{M}^{k+1}_{\text{REP}}\!\!\!\!\!\!&&=\left\{
			\begin{aligned}
				&1, \text{ if }\mathcal{M}^{k}_{\text{REP}}\!=\!1\text{ and }\mathbb{A}_{\text{sub}}[2^{k+1}\!-\!1]\!=\!1\text{ and }\xi\,, \\
				&0, \text{ otherwise} \,,
			\end{aligned} \right. \\
		\end{aligned}
	\end{small}
\end{equation}
where $\xi$ represents the common logic $\mathbb{A}_{\text{sub}}[2^{k}:2^{k+1}-2]=\mathbf{0}$ and $k\geq 1$.
The associated logic gates are shared by the identification of R0 and REP, thus reducing the overall logic area.
Subsequently, the identification logic of the \{R1, SPC, TYPE-III\} group is formulated as
\begin{equation}\label{eq:logic_r1_spc_type3}
	\begin{small}
		%\left\{
		\begin{aligned}
			&\mathcal{M}^{k+1}_{\text{R1}} &&\!\!\!\!\!\!=
			\begin{cases}
				1, & \text{if } \mathcal{M}^{k}_{\text{R1}}=1\text{ and }\zeta\,, \\
				0, & \text{otherwise} \,,
			\end{cases} \\
			&\mathcal{M}^{k+1}_{\text{SPC}} &&\!\!\!\!\!\!=
			\begin{cases}
				1, & \text{if } \mathcal{M}^{k}_{\text{SPC}}=1\text{ and }\zeta\,, \\
				0, & \text{otherwise} \,,
			\end{cases}\\
			&\mathcal{M}^{k+1}_{\text{TYPE-III}} &&\!\!\!\!\!\!=
			\begin{cases}
				1, & \text{if } \mathcal{M}^{k}_{\text{TYPE-III}}=1\text{ and }\zeta\,, \\
				0, & \text{otherwise} \,,
			\end{cases}
		\end{aligned}
		%\right.
	\end{small}
\end{equation}
where $\zeta$ represents the common logic $\mathbb{A}_{\text{sub}}[2^{k}:2^{k+1}-1]=\mathbf{1}$, $k\geq 1$ for $\mathcal{M}^k_{\text{R1}}$ and $\mathcal{M}^k_{\text{SPC}}$, and $k\geq 2$ for $\mathcal{M}^k_{\text{TYPE-III}}$ since length-$2$ TYPE-III node is invalid. 
Thus, we can identify the basic nodes at all possible lengths by recursively invoking~\eqref{eq:logic_r0_rep} and~\eqref{eq:logic_r1_spc_type3}.
Note that some preconditions are necessary to trigger the above recursive processing.
For instance,~\eqref{eq:precondition} illustrates the preconditions of the \{R1, SPC, TYPE-III\} group.\footnote{The preconditions of the \{R0, REP\} group are similar to~\eqref{eq:precondition}. Hence, we omit them in this paper for brevity.}
\begin{equation}\label{eq:precondition}
	\begin{small}
		%\left\{
		\begin{aligned}
			&\mathcal{M}^1_{\text{R1}} &&\!\!\!\!\!=
			\begin{cases}
				1, & \text{if } \mathbb{A}_{\text{sub}}[0\!:\!1]\!=\!\mathbf{1}\,, \\
				0, & \text{otherwise} \,,
			\end{cases} \\
			&\mathcal{M}^1_{\text{SPC}} &&\!\!\!\!\!=
			\begin{cases}
				1, & \text{if } \mathbb{A}_{\text{sub}}[0\!:\!1]\!=\![0,1]\,, \\
				0, & \text{otherwise} \,,
			\end{cases}\\
			&\mathcal{M}^2_{\text{TYPE-III}} &&\!\!\!\!\!=
			\begin{cases}
				1, & \text{if } \mathbb{A}_{\text{sub}}[0\!:\!3]\!=\![0,0,1,1]\,, \\
				0, & \text{otherwise} \,.
			\end{cases}
		\end{aligned}
		%\right.
	\end{small}
\end{equation}
\begin{figure}[t]
	\centering
	\includegraphics[width=1.0\linewidth]{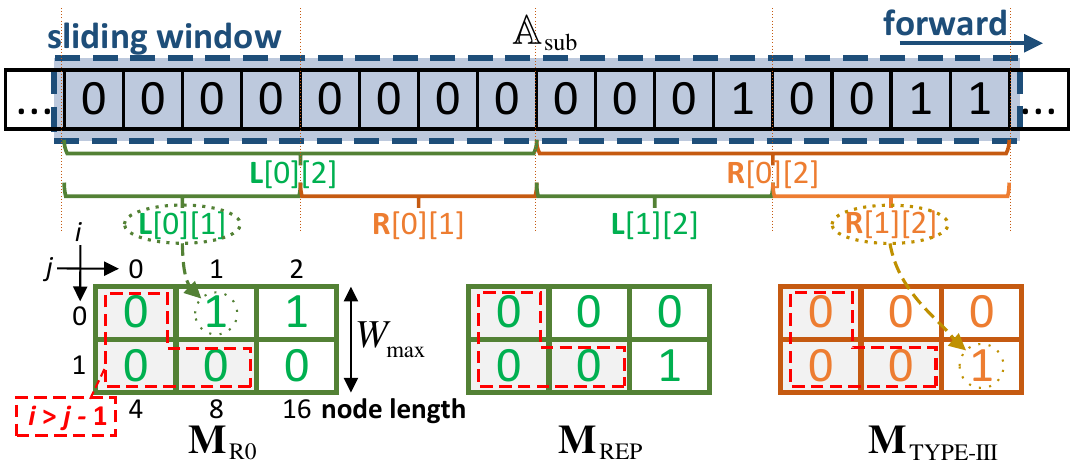}\\
	\caption{Representation of the 2D arrays describing the locations of three basic nodes (R0, REP, and TYPE-III) in the example $16$-bit sequence $\mathbb{A}_{\text{sub}}$, where the horizontal index represents the stage of the basic node $(j-i+1)$ and the vertical index represents the basic node length $(2^{j-i+1})$. The ``left'' and ``right'' segments in~$\mathbb{A}_{\text{sub}}$ are labeled green and orange, respectively.}\label{fig:Instruction_P12}
\end{figure}
\begin{figure}[t]
	\centering
	\includegraphics[width=1.0\linewidth]{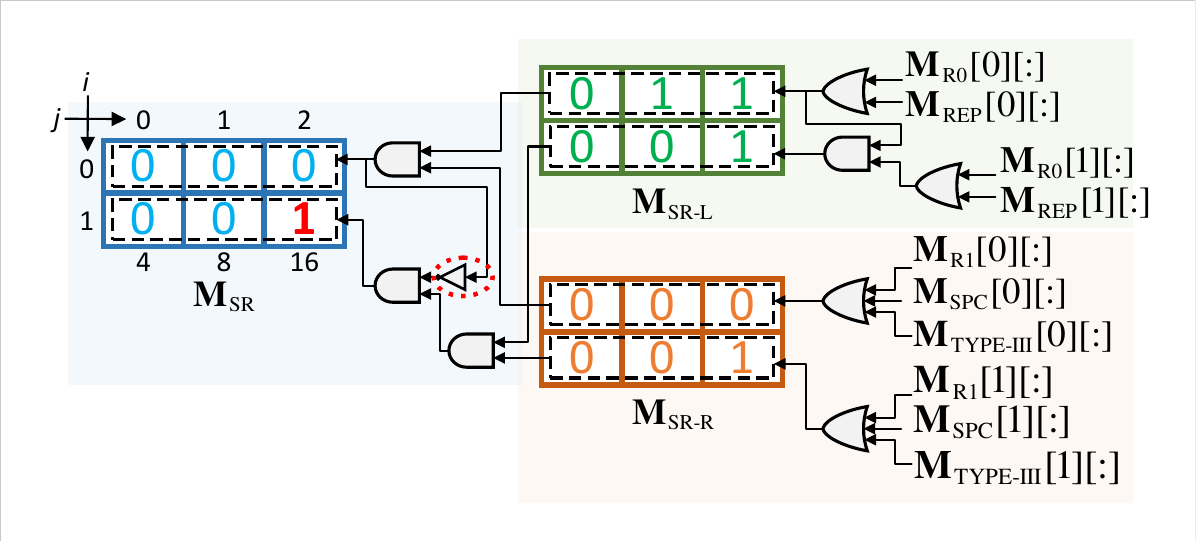}\\
	\caption{The regrouping process for SR nodes, using the 2D arrays of basic~nodes.}\label{fig:Instruction_P3}
\end{figure}

\subsection{Node Identification for SR Nodes}\label{subsec:id_sr}
As outlined in Section~\ref{subsec:nodes}, the identification of SR nodes is more complicated than that of the basic nodes since the SR node has a generalized structure and requires a recursive detection process.
To effectively tackle this challenge, we divide the detection logic for SR nodes into three phases:
\begin{enumerate}
	\setlength{\itemsep}{0pt}
	\setlength{\parsep}{0pt}
	\setlength{\parskip}{0pt}
	\item
	Detection of the ``left" low-rate part within SR nodes is responsible for identifying R0 and REP nodes located in the left regions of SR nodes at different stages.
	\item
	Detection of the ``right" high-rate part within SR nodes is responsible for identifying R1, SPC, and TYPE-III nodes as the source node in the right regions of SR nodes at different stages of the decoding tree.
	\item
	The ``regrouping'' phase prioritizes selecting the SR node with the maximum length and, as a secondary criterion, the source node of the longest feasible length.
\end{enumerate}

In Fig.~\ref{fig:Instruction_P12}, we depict an example SR node using a $16$-bit sequence $\mathbb{A}_{\text{sub}}$ in the sliding window.
In this case, we assume $N_{s_\mathrm{max}}=16$ and $W_{\mathrm{max}}=2$, where $W_{\mathrm{max}}$ denotes the maximum number of R0/REP nodes as left descendants.
We first define a two-dimensional (2D) array,~$\mathbf{M}_{\text{X}}$, per basic node type ``X'' to find the locations of all node candidates.
The dimension of this 2D is $W_{\mathrm{max}} \times (\log N_{s_\mathrm{max}}-1)$, the same for all basic node types.
The horizontal index $i$ reveals the stage of the basic node, and the vertical index $j$ reveals the basic node length.
For instance, a pair of $\{i,j\}$ represents a length-$2^{j-i+1}$ basic node that is located at the $j-i+1$-th stage, $0\leq i\leq W_{\max}-1$ and $0\leq j < \log N_{s_{\max}}-1$.
Note that each index pair, $\{i, j\}$, not only provides the position of the basic node itself, but also indicates a potential constraint of the SR node where the basic node can be located: the number of identified left-side R0/REP nodes is $i+1$, the length of the SR node is $N_s=2^{j+2}$, and the length of the source node is $N_r=2^{j-i+1}$.
It is noteworthy that such a structure of $\mathbf{M}$ (for basic nodes) does not imply that it can only be populated once the SR nodes are successfully identified.
Instead, the 2D array~$\mathbf{M}$ can be filled as long as the basic node of length-$2^{j-i+1}$ is detected under the SR constraint of each index pair $\{i,j\}$ (herein, the identification of basic nodes follows \eqrangelabel{eq:logic_r0_rep}{eq:precondition} as described in Section~\ref{subsec:id_basic}).

To facilitate the detection of the ``left'' and ``right'' parts within SR nodes, we label the segments of $\mathbb{A}_{\text{sub}}$ as $\mathbf{L}[i][j]$ or $\mathbf{R}[i][j]$, as depicted in Fig.~\ref{fig:Instruction_P12}.\footnote{Based on the SR node structure, the left segment $\mathbf{L}[i][j]$ is merely used to detect the R0 and REP nodes and the right segment $\mathbf{R}[i][j]$ only for the R1, SPC, and TYPE-III nodes.}
In phase~$1$, we check the $\mathbf{L}$ segments to identify R0 and REP nodes and then fill $\mathbf{M}_\text{R0}$ and $\mathbf{M}_\text{REP}$.
To be specific, $\mathbf{M}_\text{R0}[0][1]=1$ because a length-$4$ R0 node is found in the left segment $\mathbf{L}[0][1]$.
Similarly, in phase~$2$, we check the $\mathbf{R}$ segments to identify high-rate source nodes and populate $\mathbf{M}_\text{R1}$, $\mathbf{M}_\text{SPC}$, and $\mathbf{M}_\text{TYPE-III}$.
For example, $\mathbf{M}_\text{TYPE-III}[1][2]=1$ because a length-$4$ TYPE-III node is found in the right segment $\mathbf{R}[1][2]$.
The arrays of R1 and SPC nodes can be generated similarly, yet, with all zeros in this example.
Note that the red dashed rectangle on a gray background represents a zone without an SR node structure.
This area is directly marked as $0$, reflecting that the source node is at its minimum length and that~$N_s$~must~exceed~$N_r$.
\begin{figure}[t]
	\centering
	\includegraphics[width=1.0\linewidth]{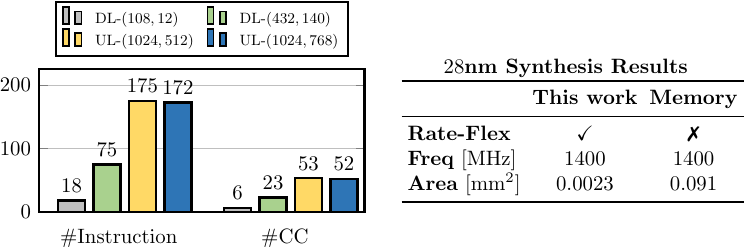}\\
	\caption{Implementation results of our instruction generator.}\label{fig:instr}
\end{figure}

For phase~$3$, ``regrouping'' is introduced in Fig.~\ref{fig:Instruction_P3} to extract the longest SR node that incorporates the source node of maximum length.
Herein, three new 2D arrays, $\mathbf{M}_{\text{SR-L}}$, $\mathbf{M}_{\text{SR-R}}$, and $\mathbf{M}_{\text{SR}}$, are constructed, by aggregating the $\mathbf{M}$ of basic nodes through simple combinatorial logic.
Note that the entire SR node detection is from the top down.
For $\mathbf{M}_{\text{SR-L}}$, the process begins with a straightforward bit-wise OR operation for $\mathbf{M}_{\text{R0}}[0][:]$ and $\mathbf{M}_{\text{REP}}[0][:]$ at the horizontal index $i=0$.
For the subsequent stage, $i=1$, an additional AND gate with the vector from the previous stage is performed.
This step ensures that a left-part low-rate node at the current stage is considered in the SR node only if R0/REP nodes have been identified at all previous source node stages.
This recursive approach can continue if $W_{\max}$ exceeds $2$.
On the other hand, $\mathbf{M}_{\text{SR-R}}$ is formed through a simple bit-wise OR operation among $\mathbf{M}_{\text{R1}}$, $\mathbf{M}_{\text{SPC}}$, and $\mathbf{M}_{\text{TYPE-III}}$.
The final step involves $\mathbf{M}_{\text{SR}}$, which searches positions where both $\mathbf{M}_{\text{SR-L}}$ and $\mathbf{M}_{\text{SR-R}}$ are marked with ``$1$'', indicating the presence of SR nodes.
Notably, the NOT gate highlighted by a red dashed circle reveals that an SR node with a shorter source node will be activated only if no source node of greater length is available.
In this case, we obtain the only SR node in $\mathbf{M}_{\text{SR}}[1][2]=1$, based on which, we detect a length-$16$ SR node with a length-$4$ TYPE-III node as its source node, corresponding to the $\mathcal{N}_{4,1}$ depicted in Fig.~\ref{fig:TREE}.
If more than one valid SR node is found in $\mathbf{M}_{\text{SR}}$, we give priority to the one with the longest length (i.e., the upper right corner of the array $\mathbf{M}_{\text{SR}}$ enjoys~a~higher~priority).

\subsection{Hardware Results of Online Instruction Generator}\label{subsec:result_inst}
While operating continuously, our proposed online instruction generator can identify one node and generate the corresponding series of instructions regarding this node in each CC.
For each node ultimately identified (only if the SR structure does not exist, we then consider basic node types), at least two instructions are generated: one for the SCU and one for the NPU. 
As illustrated in Fig.~\ref{fig:instr}, our instruction generator only requires $53$ CCs (i.e., one node per cycle) to complete the generation of 175 instructions required for UL-$(1024, 512)$ (the worst-case in UL).
To prevent overloading the decoder cores with a considerable number of instructions at multiple CCs, which would exceed its processing capacity, we have integrated a compact FIFO buffer between the decoder cores and the controller, as shown in Fig.~\ref{fig:top}.
This buffer temporarily holds the newly generated instructions, pausing further node identification when the FIFO fills up.
To highlight the advantages of our online instruction generator, we compared it with a straightforward instruction memory\cite{Ren22Sequence} that holds instructions generated offline, providing a capacity of $175$ instructions (i.e., the demand for UL-$(1024, 512)$).
Given that two interleaved frames are decoded simultaneously, the required memory capacity further doubles.
Synthesis results in $28$nm FD-SOI technology show our online generator occupies only an area of $0.0023$ mm$^2$.
This area is $97.5\%$ smaller and enjoys a comparable data path length compared to straightforward instruction memory.
Crucially, our approach supports on-the-fly instruction generation, enabling decoder rate flexibility.
\section{Implementation Results}\label{sec:implem}
\begin{table}[t]
	\centering
	\caption{Implementation results of our decoders$^\dagger$.}
	\label{tab:hw_my_results}
	
	\scriptsize
	\tabcolsep 0.7mm
	\def\CmidW{0.08cm}
	
	\resizebox{\columnwidth}{!}{\begin{tabular}{lcccccccc}
			\toprule
			~ & \mc{4}{c}{\textbf{This work}$^\ddagger$} & \mc{4}{c}{TSP'22~\cite{Ren22Sequence}} \\
			\cmidrule(l{\CmidW}){2-5} \cmidrule(l{\CmidW}r{\CmidW}){6-9}
			\textbf{Length [bit]} & \mc{2}{c}{$512$} & \mc{2}{c}{$1024$} & \mc{2}{c}{$512$} & \mc{2}{c}{$1024$} \\
			\textbf{List-size}         & $4$   & $8$    & $4$   & $8$   & $4$    & $8$   & $4$   & $8$   \\
			\textbf{CCs}              & $76$   & $87$    & $188$   & $212$   & $155$    & $173$   & $355$   & $395$   \\
			%\cmidrule(l{\CmidW}){2-3} \cmidrule(l{\CmidW}r{\CmidW}){4-5} \cmidrule(l{\CmidW}){6-7} \cmidrule(l{\CmidW}){8-9}
			\textbf{Latency} [$\mathrm{\mu s}$]      & $0.064$  & $0.089$   & $0.161$  & $0.22$  & $0.119$   & $0.174$  & $0.28$   & $0.40$  \\
			\textbf{Area} [mm${^2}$]  & $0.327$ & $0.648$  & $0.425$ & $0.838$ & $0.211$  & $0.439$ & $0.286$ & $0.608$ \\
			\textbf{Freq.} [MHz]         & $1181$  & $970$   & $1168$  & $965$  & $1302$   & $994$  & $1255$  & $977$    \\
			\cmidrule(l{\CmidW}){2-5} \cmidrule(l{\CmidW}r{\CmidW}){6-9}
			\textbf{Coded T/P} \![Gbps]           & $7.956$ & $5.709$  & $6.362$ & $4.661$ & $4.301$  & $2.942$ & $3.619$ & $2.532$ \\
			\textbf{T/P gain}          & $\uparrow\!85\%$ & $\uparrow\!\bm{94\%}$  & $\uparrow\!76\%$ & $\uparrow\!\bm{84\%}$ & $-$  & $-$ & $-$ & $-$ \\
			\cmidrule(l{\CmidW}){2-5} \cmidrule(l{\CmidW}r{\CmidW}){6-9}
			\parbox{1.8cm}{\raggedright \textbf{Area Eff.}\\ {[Gbps/mm$^2$]} } & $24.33$ & $8.81$ & $14.97$ & $5.56$ & $20.43$ & $6.704$ & $12.67$ & $4.165$ \\
			\textbf{Area Eff. gain}$^\dagger$           & $\uparrow\!19\%$ & $\uparrow\!\bm{32\%}$  & $\uparrow\!18\%$ & $\uparrow\!\bm{34\%}$ & $-$  & $-$ & $-$ & $-$ \\
			\bottomrule
	\end{tabular}}
	\begin{tablenotes}
		\footnotesize
		\item[*] $^\dagger$ While our works are compatible with all DL and UL codes, latency is obtained at DL-$(432,140)$ and UL-$(1024,512)$ codes for length-$512$ and length-$1024$ decoders, respectively.
		\item[*] $^\ddagger$ We set the decoder to Mode-III for enhanced average T/P and reliability.
	\end{tablenotes}
\end{table}

In this section, we present the implementation results of our generalized node-based SCL decoder with frame-interleaving, based on a STM 28nm FD-SOI technology. 
The decoder is synthesized with Synopsys Design Compiler and placed and routed using Cadence Innovus. Power analysis is done under typical operating conditions with parasitic and activity back annotation ($1.0$~V and $25^\circ$C). 
We follow the default configurations of~\cite{Ren22Sequence} to fix the number of SCU stages ($\#$SCU) to $3$, the number of PEs at the first stage of the SCU ($\#$PE) to $64$, and the maximum size of nodes to $N_{s_{\max}}=32$.
We select $W_{\mathrm{max}}=2$ as the best trade-off between decoding time and area (i.e., node identification of Section~\ref{sec:instruction} can identify the SR nodes with at most two R0/REP left descendants). 
Moreover, as {length-$32$} SR nodes are rarely found in 5G~NR polar codes, we constrain the RSU size to $16$. 
To strike a balance between error-correcting performance and hardware complexity, as shown in Fig.~\ref{fig:FER}, we employ a sign-magnitude quantization with the $6$-bit LLRs and $7$-bit PMs.
This choice incurs a negligible FER performance loss compared to floating point. 
Note that we present two versions of our decoder, one tailored to DL with a maximum code length of $512$ and one for UL with a maximum code length of $1024$ bits.
\begin{figure}[t]
	\centering
	\begin{minipage}{.475\linewidth}
		\centering
		\includegraphics[width=0.85\columnwidth]{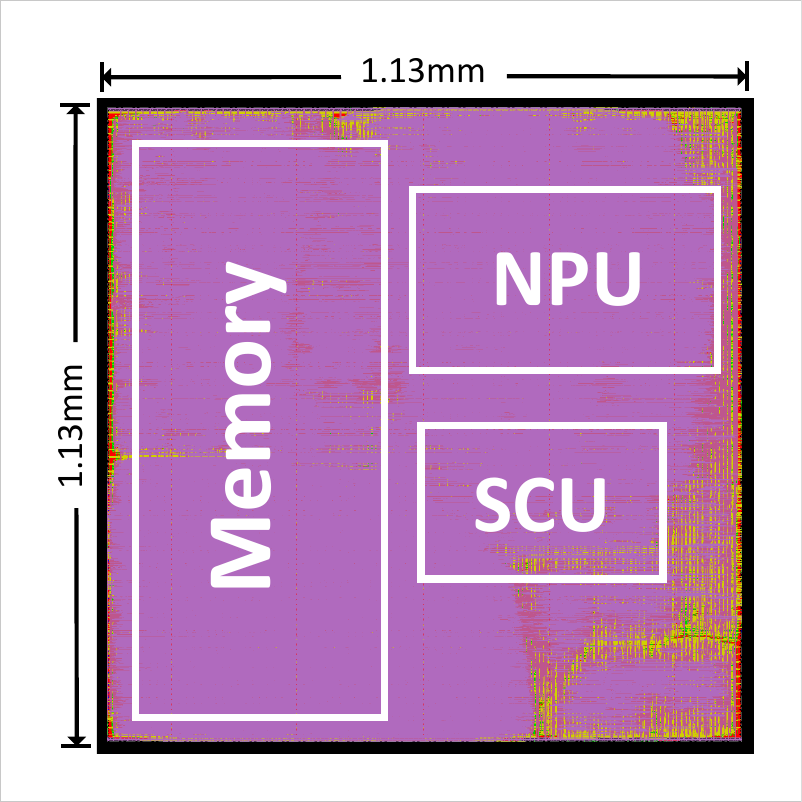}
	\end{minipage}
	\begin{minipage}{.475\linewidth}
		\centering
		\scriptsize
		\tabcolsep 1mm
		\renewcommand{\arraystretch}{1.125}
		\begin{tabular}{lc}
			\toprule
			\textbf{Technology}  & $28$nm FD-SOI \\ \midrule
			\multicolumn{1}{l|}{\textbf{Quantization} [bit]} & $6$ \\
			\multicolumn{1}{l|}{\textbf{Core Area} [mm$^{2}$]} & $1.13\!\times\!\!1.13$ \\
			\multicolumn{1}{l|}{\textbf{Gate Count} [M]} & $1.45$ \\ % 1274493/(1.41*0.41: C12T28SOI_LR_NAND2X5_P0)
			\multicolumn{1}{l|}{\textbf{Voltage} [V]} & $1.0$ \\
			\multicolumn{1}{l|}{\textbf{Frequency} [MHz]} & $692$ \\
			\multicolumn{1}{l|}{\textbf{Coded T/P} \![Gbps]} & $3.34$ \\
			\multicolumn{1}{l|}{\textbf{Power} [mW]} & $201.6$ \\
			\multicolumn{1}{l|}{\textbf{Energy} [pJ/bit]} & $60.32$ \\
			\bottomrule
		\end{tabular}
	\end{minipage}
	\caption{A post-layout of our length-$1024$ SCL decoder with $L=8$ implemented in a $28$nm process, wherein the white boxes represent the memory and two computational units, the SCU and the NPU.}
	\label{fig:SecV_postlayout}
\end{figure}

\begin{table*}[]
	\scriptsize
	\tabcolsep 1.875mm
	\renewcommand{\arraystretch}{1.05}
	\def\CmidW{0.08cm}
	\caption{Comparisons with the state-of-the-art SCL decoders for the UL-$(1024,512)$ code.}
	\label{tab:5Gresults}
	\centering
	\begin{tabular}{lccccccccccc}
		\toprule
		& \multicolumn{2}{c}{\multirow{2}{*}{\textbf{This work}}} & JETCAS'17 & ISWCS'18 & JSSC'20 & VLSI'22 & TCAS-I'23 & ISTC'23$^{\ddagger}$ & TSP'15 & TSP'17 & TSP'22  \\
		&    & & \cite{giard2017polarbear} & \cite{liu20185} & \cite{tao2020configurable} & \cite{Kam22Cost} & \cite{Kam23low} & \cite{johannsen2023successive} & \cite{Bala15LLR} & \cite{Hashemi17Fast} & \cite{Ren22Sequence} \\ \cmidrule(l{\CmidW}){2-3}  \cmidrule(l{\CmidW}){4-4} \cmidrule(l{\CmidW}){5-5}
		\cmidrule(l{\CmidW}){6-6} \cmidrule(l{\CmidW}){7-7}
		\cmidrule(l{\CmidW}){8-8} \cmidrule(l{\CmidW}){9-9}
		\cmidrule(l{\CmidW}){10-10} \cmidrule(l{\CmidW}){11-11} \cmidrule(l{\CmidW}){12-12}
		\textbf{Technology} {[}nm{]}         & \mc{2}{c}{$28$}             & $28$         & $16$           & $40$         & $28$      & $65$  & $12$        & $90$        & $65$      & $28$                    \\
		\textbf{Implementation}              & Synthesis            & Post-layout     & Silicon    & Silicon      & Silicon  & Silicon & Post-layout   & Post-layout  & Synthesis   & Synthesis &  Synthesis          \\
		\textbf{Full 5G~NR} & \mc{2}{c}{$\checkmark$} & \ding{55} & $\checkmark$ & $\checkmark$ & $\checkmark$ & $\checkmark$ & \ding{55} & \ding{55} & \ding{55} & $\checkmark$ \\
		\textbf{Permutations}$^\diamond$ & \mc{2}{c}{$\checkmark$} & \ding{55} & \ding{55} & \ding{55} & \ding{55} & \ding{55} & $\checkmark$  & \ding{55} & \ding{55} & \ding{55} \\
		\textbf{Rate-Flexible} & \mc{2}{c}{$\checkmark$} & $\checkmark$ & $\checkmark$ & $\checkmark$ & \ding{55} & \ding{55} & \ding{55} & $\checkmark$ & \ding{55} & \ding{55}  \\
		\textbf{Voltage} {[}V{]}             & \mc{2}{c}{$1.0$}        & $1.3$          & $0.9$            & $0.9$        & $1.05$       & $1.1$   & $0.8$     & $1.0$         & $-$     & $1.0$                \\
		\textbf{List-size}                 & \mc{2}{c}{$8$}                    & $4$         & $8$            & $2$          & $8$      & $8$     & $8$    & $8$           & $8$      & $8$                 \\
		\textbf{CCs} & \mc{2}{c}{$212$}           & $2408$         & $790$            & $136$          & $454$      & $304$     & $-$    & $2662$           & $618$      & $395$               \\
		\textbf{SNR$@\text{FER}=10^{-3}$} & \mc{2}{c}{$1.98$}  & $2.3$ & $2.1$ & $2.63$ & $2.1$ & $2.1$ & $3.05$ & $2.1$ & $2.1$ & $2.1$ \\
		\textbf{Latency} [$\mathrm{\mu s}$]          & $0.22$             & $0.306$         & $3.34$            & $0.72$          & $0.32$      & $1.10$         & $1.003$    & $0.064$       & $4.18$      & $0.85$        & $0.40$              \\
		\textbf{Area} {[}mm$^{2}${]}           & $0.838$           & $1.28$      & $0.44$         & $0.06$       & $0.637$    & $0.595$       & $3.89$   & $0.381$     & $3.58$    & $3.975$       & $0.608$           \\
		\textbf{Frequency} {[}MHz{]}         & $965$             & $692$      & $721$          & $1100$        & $430$     & $413$      & $300$   & $500$      & $637$     & $722$       & $977$            \\
		\textbf{Coded T/P} {[}Gbps{]}              & $4.661$            & $3.34$      & $0.307$         & $1.426$       & $3.25$     & $0.925$        & $1.02$  & $64.0$     & $0.246$   & $1.198$      & $2.532$           \\
		\textbf{Power} {[}mW{]}              & $-$            & $201.6$       & $128.3$          & $-$      & $42.8$   & $101.4$       & $389$   & $552$      & $-$   & $-$     & $-$            \\ \midrule
		
		\multicolumn{11}{l}{Scaled to $28$nm, $1.0$~V$^{\dagger}$}\\
		\textbf{Area} {[}mm$^{2}${]}           & $0.838$      & $1.28$      & $0.44$         & $0.184$      & $0.312$  & $0.595$      & $0.722$   & $2.074$      & $0.347$   & $0.738$     & $0.608$    \\
		\textbf{Coded T/P} {[}Gbps{]}              & $4.661$      & $3.34$       & $0.307$         & $0.815$       & $4.643$  & $0.925$      & $2.368$   & $27.43$     & $0.791$    & $2.781$     & $2.532$     \\
		\textbf{Area Eff.} {[}Gbps/mm$^{2}${]} & $5.56$       & $2.62$      & $0.692$         & $4.435$       & $14.87$  & $1.56$      & $3.28$   & $13.24$     & $2.282$   & $3.77$     & $4.165$     \\
		%\textbf{Power} {[}mW{]} & $-$  & $201.6$  &  $75.92$  &  $-$  &  $36.99$  &  $91.97$  &  $138.5$  &  $2012.5$ & $-$  &  $-$  &  $-$ \\
		%\textbf{Energy} {[}pJ/bit{]}    & $-$       &   $60.32$           &   $247.3$             &   $-$          &     $7.97$      &  $99.43$            &    $58.49$   &    $73.37$        &  $-$         &    $-$         &     $-$       \\
		\textbf{Power} {[}mW{]} & $-$  & $201.6$  &  $75.92$  &  $-$  &  $52.84$  &  $91.97$  &  $321.49$  &  $862.5$ & $-$  &  $-$  &  $-$ \\
		\textbf{Energy} {[}pJ/bit{]}    & $-$       &   $60.32$           &   $247.3$             &   $-$          &     $11.38$      &  $99.43$            &    $135.77$   &    $31.44$        &  $-$         &    $-$         &     $-$       \\						
		\bottomrule
	\end{tabular}
	\begin{tablenotes}
		\item[*] $^{\dagger}$ Scaled to $28$nm and $1.0$~V with area $\varpropto \sigma^2$, frequency $\varpropto 1/\sigma$, and power $\varpropto 1/u^{2}$, where $\sigma$ is the scaling factor to $28$nm and $u$ is the scaling factor to $1.0$~V.
		\item[*] $^{\ddagger}$ Only~\cite{johannsen2023successive} is implemented for $(128, 60)$ codes while the others are for $(1024, 512)$ polar codes.
		\item[*] $^{\diamond}$ Our work supports graph ensemble decoding and~\cite{johannsen2023successive} uses automorphisms to enhance error-rate performance.
	\end{tablenotes}
\end{table*}

\subsection{Implementation Results of Our Decoder}
Table~\ref{tab:hw_my_results} summarizes the synthesis results of our proposed decoder tailored for DL and UL polar codes with $L\in\{4,8\}$. 
While our decoders are compatible with all 5G~NR polar codes, the latency of Table~\ref{tab:hw_my_results} focuses on two classical codes: DL-$(432, 140)$ and UL-$(1024, 512)$.
In comparison with the baseline decoder~\cite{Ren22Sequence}, our length-$512$ decoder with $L=8$ yields a $47.6\%$ area increase (attributed to two memory instances for interleaving) yet significantly reduces the worst-case latency from $173$ cycles to $87$ cycles (nearly half).
Moreover, since the various working modes (see Section~\ref{sec:hardware}) and the dynamic strategies to eliminate stalls (see Section~\ref{sec:strategy}) are only involved with the control signal interactions, our work shares a comparable critical path and a similar maximum frequency with the baseline decoder.
Consequently, our length-$512$ decoder with $L=8$ attains a throughput of~$5.709$~Gbps and an area efficiency of~$8.81$~Gbps/mm$^2$, surpassing~\cite{Ren22Sequence} by $94\%$ and $32\%$, respectively.
For length-$1024$ codes at half rate, while our decoder has a $37.8\%$ increase in cell area to $0.838$~mm$^{2}$, it delivers an $84\%$ increase to $4.661$~Gbps in throughput and improves area efficiency by $34\%$ to $5.56$~Gbps/mm$^2$.

Fig.~\ref{fig:SecV_postlayout} depicts the post-layout results of our length-$1024$ SCL-$8$ decoder.
When all the physical design steps are completed, the post-layout of our decoder reveals a core size of $1.13\times 1.13$~mm$^2$ with a cell utilization of $65.6\%$.
Due to the effects of the physical wire lengths and parasitic delays, the maximum operating frequency is reduced to $692$~MHz.
When decoding UL-$(1024, 512)$ code with a throughput of $3.34$~Gbps, this decoder has a dynamic power consumption of $201.6$~mW and an energy consumption of $60.32$~pJ/bit.

\subsection{Comparison with Previous Works}
Table~\ref{tab:5Gresults} provides a comprehensive comparison between our frame-interleaving decoder and the state-of-the-art polar decoder implementations in~\cite{giard2017polarbear,liu20185,tao2020configurable,Kam22Cost,Kam23low,johannsen2023successive,Bala15LLR,Hashemi17Fast,Ren22Sequence}.
To ensure fairness, we normalize all previous works to a $28$~nm process with a supply voltage of $1.0$~V.\footnote{The technology scaling method refers to Table~I of~\cite{stillmaker2017scaling}. Note that this is rough and optimistic since it does not apply to the interconnects~\cite{yin2017high, mo202112}.}
Compared to a similar node-based SCL decoder with the same $28$~nm technology~\cite{Kam22Cost}, our work has a $3.61\times$ throughput, $1.68\times$ area efficiency, and $39.3\%$ less energy consumption. 
When compared to conventional SCL decoder without node-based techniques~\cite{giard2017polarbear,tao2020configurable}, our decoder enjoys a throughput that is $10.89\times$ faster than~\cite{giard2017polarbear} and $4.10\times$ higher than~\cite{liu20185}.
Compared to the SCL decoder with overlapped operation units in~\cite{Kam23low} that leverages parallel processing, our work has a $30.2\%$ reduction of decoding latency to only $212$~CCs per frame when decoding the UL-$(1024, 512)$ code.
Although our area efficiency is $20.1\%$ inferior to that of~\cite{Kam23low}, we present a $41.1\%$ higher throughput and achieve a $55.6\%$ lower energy consumption.
In addition, our decoder supports graph ensemble decoding (in Mode-II and Mode-III), thus improving the error-correcting performance by $0.18$~dB at an FER of~$10^{-3}$ (identical to SCL-$16$ decoding shown in Fig.~\ref{fig:FER}).  
While our area efficiency and energy efficiency are inferior to the SC automorphism list (SCAL) decoder in~\cite{johannsen2023successive}, given that~\cite{johannsen2023successive} is mainly restricted to short-length codes and hard-wired routing networks, our decoder is fully compatible with all 5G~NR polar codes and supports a flexible permutation generator.
Compared to~\cite{johannsen2023successive}, our decoder has a $76.6\%$ lower power consumption.
It is noteworthy that the integrated online instruction generator contributes to the rate-flexibility of our decoder, which is not achieved by the previous node-based works~\cite{Kam22Cost, Kam23low, Hashemi17Fast, Ren22Sequence} mentioned~above.
Moreover, our proposed decoder supports two frames/graphs decoding simultaneously, given that $2\times$ can achieve reasonable improvements for 5G scenarios. 
This work can also be extended to multiple cores for higher~parallelism.

\section{Conclusion}\label{sec:conclusion}
The architecture of frame-interleaving significantly enhances the efficiency of node-based SCL decoders by eliminating processing delays from data-dependent computational units.
This improvement is achieved by reusing otherwise idle processing units to decode two frames simultaneously. 
However, simple interleaving retains some residual stalls.
Our dynamic stall-reduction strategies effectively remove these stalls by reasonably reorganizing the decoding schedule, allowing for considerable overlapped operations without architectural modifications.
Moreover, our frame-interleaving decoder supports graph ensemble decoding and various modes of operation aiming at higher throughput, performance, and efficiency.
We also introduce an online instruction generator, tailored to SR nodes, that ensures low decoding latency and rate flexibility in our generalized node-based SCL decoder.
The corresponding $28$nm FD-SOI ASIC decoder with $L=8$ demonstrates that our work has a throughput of $3.34$~Gbps and an area efficiency of $2.62$~Gbps/mm$^2$ for the UL-$(1024,512)$ code.

\bibliographystyle{IEEEtran}
\bibliography{IEEEabrv,mybib}

\end{document}